%% file: neurips_2025.tex
\documentclass{article}

\usepackage[numbers]{natbib}
\usepackage[final]{neurips_2025}

\usepackage[utf8]{inputenc} 
\usepackage[T1]{fontenc}    
\usepackage{hyperref}       
\usepackage{url}            
\usepackage{booktabs}       
\usepackage{amsfonts}       
\usepackage{nicefrac}       
\usepackage{microtype}      
\usepackage{xcolor}         

\input{preamble}

\title{Reverse Diffusion Sequential Monte Carlo Samplers}

\author{%
Luhuan Wu\thanks{Correspondence email: \texttt{lw2827@columbia.edu}} \\
  Columbia University \\
  \And
   Yi Han \\
   Columbia University \\
   \AND
   Christian A. Naesseth \\
  University of Amsterdam
  \And
  John P. Cunningham \\
  Columbia University \\
}

\begin{document}

\maketitle

\input{sections/abstract}

\input{sections/introduction}

\input{sections/background}

\input{sections/method}

\input{sections/related_works}

\input{sections/experiments}

\input{sections/discussion}

\bibliographystyle{unsrtnat}
\bibliography{reference}

\newpage
\input{sections/appendix}



\end{document}

%% file: preamble.tex
\usepackage{amsmath, amssymb, amsfonts, amsthm, mathtools}
\usepackage{graphicx}
\usepackage[linesnumbered,ruled,vlined]{algorithm2e}
\usepackage{multirow}
\usepackage[shortlabels]{enumitem}
\usepackage{graphicx}
\usepackage{subcaption}
\usepackage{natbib}
\usepackage{cancel}
\usepackage[sort,compress]{cleveref}
\creflabelformat{equation}{(#2#1#3)}
\crefname{equation}{eq.}{eqs.}
\Crefname{equation}{Eq.}{Eqs.}
\Crefname{section}{\S}{\S}
\crefname{enumi}{Step}{Steps}

\usepackage{makecell} 

\SetKwInput{KwInput}{Input}
\SetKwInput{KwOutput}{Output}
\crefname{algocf}{algorithm}{algorithms}
\Crefname{algocf}{Algorithm}{Algorithms}

\newcommand{\calD}{\mathcal{D}}

\newtheorem{theorem}{Theorem}

\newtheorem{assumption}{Assumption}

\crefname{assumption}{assumption}{assumptions}
\Crefname{assumption}{Assumption}{Assumptions}
\crefname{definition}{definition}{definitions}
\Crefname{definition}{Definition}{Definitions}
\crefname{proposition}{proposition}{propositions}
\Crefname{proposition}{Proposition}{Propositions}

\newcommand{\BlackBox}{\rule{1.5ex}{1.5ex}}
\renewenvironment{proof}{\par\noindent{\bf Proof }}{\hfill\BlackBox\newline}

\newcommand{\xt}{x_{t}}
\newcommand{\xtp}{x_{t+1}}
\newcommand{\xtm}{x_{t-1}}
\newcommand{\xT}{x_{T}}
\newcommand{\xzero}{x_{0}}
\newcommand{\xone}{x_{1}}
\newcommand{\xn}[1]{x_{#1}}

\definecolor{auxcolor}{RGB}{80,120,160}  

\newcommand{\ut}{\textcolor{auxcolor}{u_t}}
\newcommand{\utp}{\textcolor{auxcolor}{u_{t+1}}}

\newcommand{\uT}{\textcolor{auxcolor}{u_{T}}}
\newcommand{\uzero}{\textcolor{auxcolor}{u_{0}}}
\newcommand{\un}[1]{\textcolor{auxcolor}{u_{#1}}}
\newcommand{\utpn}[1]{\textcolor{auxcolor}{u_{t+1}^{#1}}}
\newcommand{\utn}[1]{\textcolor{auxcolor}{u_{t}^{#1}}}

\newcommand{\utsupm}{\textcolor{auxcolor}{u_t^{(m)}}}
\newcommand{\vi}
{\textcolor{auxcolor}{v_i}}
\newcommand{\nui}
{\textcolor{auxcolor}{v_i}}

\newcommand{\ft}{f_{t}}
\newcommand{\ftp}{f_{t+1}}

\newcommand{\gt}{g_{t}}
\newcommand{\gtp}{g_{t+1}}

\newcommand{\wt}{w_{t}}
\newcommand{\wtp}{w_{t+1}}

\newcommand{\wT}{w_{T}}

\newcommand{\ourmethod}{RDSMC}
\newcommand{\ourmethodfull}{Reverse Diffusion Sequential Monte Carlo}

\newcommand{\tstart}{ t_\mathrm{start\text{-}resampling}}
\newcommand{\essthreshold}{\kappa_{\mathrm{\text{ess}}}}
\newcommand{\nsteps}{n} 
\newcommand{\msteps}{l} 
\newcommand{\stepsize}{\delta_{\text{mcmc}}}
\newcommand{\nis}{M}

\newcommand{\tinit}{t_\mathrm{\text{init}}}
\newcommand{\tfinal}{t_\mathrm{\text{final}}}

\newcommand{\fric}{\delta_{\gamma}}

\newcommand{\stepsizeinit}{\delta_{\text{mcmc-init}}}

\newcommand{\bigO}{\mathcal{O}}

\newtheoremstyle{italicheader}
  {3pt}   
  {3pt}   
  {}      
  {}      
  {\itshape} 
  {.}     
  {0.5em} 
  {}      

\theoremstyle{italicheader}
\newtheorem{observation}{Observation}
\crefname{observation}{Observation}{Observations}

%% file: sections/abstract.tex
\begin{abstract}
We propose a novel sequential Monte Carlo (SMC) method for sampling from unnormalized target distributions based on a reverse denoising diffusion process. 
While recent diffusion-based samplers simulate the reverse diffusion using approximate score functions, they can suffer from accumulating errors due to time discretization and imperfect score estimation. 
In this work, we introduce a principled SMC framework that formalizes diffusion-based samplers as proposals while systematically correcting for their biases. 
The core idea is to construct informative intermediate target distributions that progressively steer the sampling trajectory toward the final target distribution. 
Although ideal intermediate targets are intractable, we develop \emph{exact approximations} using quantities from the score estimation-based proposal, without requiring additional model training or inference overhead. 
The resulting sampler, termed \textit{\ourmethodfull}, enables consistent sampling and unbiased estimation of the target's normalization constant under mild conditions. 
We demonstrate the effectiveness of our method on a range of synthetic targets and real-world Bayesian inference problems. \footnote{Our code is available at \url{https://github.com/LuhuanWu/RDSMC}.} 
\end{abstract}

%% file: sections/introduction.tex
\section{Introduction}

Sampling from unnormalized target distributions is a fundamental problem in many applications, ranging from Bayesian inference \citep{andrieu2003introduction,robert2004metropolis} to simulating molecular systems \citep{frenkel2000molecular}. Classical methods like  Markov chain Monte Carlo (MCMC) simulate a Markov chain with the target as its stationary distribution, but they can suffer from slow mixing and difficulty traversing between modes for complex distributions. 
Particle methods such as importance sampling generate exact samples in a large compute limit; yet they struggle with the curse of dimensionality \citep{chatterjee2018sample}. Alternatively, variational inference (VI) \citep{blei2017variational} casts inference as an optimization task, though its success depends on the expressiveness of the variational family and the complexity of the optimization landscape.

Recently, diffusion models have emerged as a powerful approach for sampling from complex distributions \citep{ho2020denoising,song2020score}.  
They define a forward noising process that gradually transforms a complex target distribution into a simple base distribution. A reverse denoising process then reconstructs  target samples by simulating the dynamics backward in time, starting from the base distribution and guided by a time-dependent score function. In the generative modeling setting, this score function is approximated by a neural network trained on samples from the target distribution. However, in the sampling context,  such training data are unavailable and only an unnormalized target density is accessible. 

Recent works on diffusion-based samplers explore alternative ways to approximate the score function directly from the target density, enabling sampling without access to training data. One line of works, known as  \textit{diffusion Monte Carlo (MC) samplers}, estimates the score function using MC methods. 
\citet{huang2023reverse,grenioux2024stochastic}  consider Langevin-style MC algorithms, but they rely on a good initialization of the reverse diffusion process to ensure theoretical guarantees. \citet{he2024zeroth} propose an alternative scheme based on rejection sampling which relaxes prior assumptions and improves sampling efficiency in low-dimensional regimes. These approaches demonstrate both theoretical and empirical advantages over conventional MCMC methods, particularly for multi-modal distributions. 

In contrast to relying on MC estimation during sampling time, a complementary line of research trains a neural network in advance to approximate the score function. To this end, some works propose new score matching objectives \citep{phillips2024particle, akhound2024iterated,ouyang2024bnem}, while variational approaches instead optimize  divergences between forward and reverse diffusion processes  \citep{zhang2021path, berner2022optimal, richter2023improved,vargas2023denoising,vargas2023transport}. 

While promising, diffusion-based samplers suffer from two sources of bias: discretization error in simulating the reverse diffusion process and approximation error in the estimated or learned score function. An exception is \citet{phillips2024particle}  which mitigates the bias of a \textit{trained} diffusion-based sampler using Sequential Monte Carlo (SMC), a general  inference tool for sequential models \citep{naesseth2019elements,chopin2020introduction}.  However, such training-based methods remain computationally complex compared to classical sampling methods and often rely on special neural network preconditioning \citep{he2025no}.

To address these challenges, we develop a new  diffusion-based sampler, \textit{\ourmethodfull} (\ourmethod) for  sampling from unnormalized target distributions, which is training-free and admits theoretical guarantees. 
Inspired by prior work, we formalize diffusion MC samplers as proposal mechanisms within an SMC framework. At a high level, \ourmethod{} generates multiple \textit{particles} from the reverse diffusion dynamics using MC-based score estimates. To correct the bias in proposals, we introduce intermediate target distributions that guide resampling of particles at each step, progressively steering them toward the final target distribution of interest. 

Crucially, our intermediate targets are efficiently computed using byproducts of MC-based score estimates, incurring no additional cost. 
Moreover, they form an \textit{exact approximation} to the ideal intermediate targets that maximize sampling efficiency, defined by the marginal distributions of an extended final target \citep{naesseth2019elements,andrieu2010particle}. This design helps particles stay closely aligned with the final target throughout the sampling process.  In contrast,  \citet{phillips2024particle} learn neural network-based surrogates that require additional training and attain the ideal target only at the final step.

\ourmethod{} belongs to a class of nested SMC methods \citep{naesseth2015nested,naesseth2019high}, inheriting the standard SMC guarantees. In particular, it produces asymptotically exact samples from the target   in the limit of many particles, and provides an unbiased estimate of the normalization constant for any fixed size of particles.

Our contributions are summarized as follows: 
\begin{itemize}
    \item We propose a new SMC algorithm, \ourmethodfull{}  (\ourmethod{}), based on the reverse diffusion process for sampling from unnormalized distributions. 
    \item \ourmethod{} is training-free and extends existing diffusion MC samplers to achieve asymptotically exact sampling and provide unbiased estimates of the normalization constant, while incurring almost no computational overhead given the same number of final samples.
    \item  Empirically, \ourmethod{} outperforms or matches existing diffusion MC samplers and classical geometric annealing-based Annealed Importance Sampling (AIS) \citep{neal2001annealed} and SMC samplers \citep{del2006sequential}
    on  synthetic targets and Bayesian logistic regression benchmarks. 
\end{itemize}

%% file: sections/background.tex
\section{Background}
\paragraph{Diffusion models.}
Diffusion models \citep{ho2020denoising,song2020score} 
 evolve a complex target distribution $\pi(x)$ into a simple base distribution $\pi_1(x)$, e.g. $\pi_1(x)=\mathcal{N}(0,1)$, via  a forward stochastic differential equation (SDE) 
\begin{align}\label{eq:forward-sde}
    d\xt = f(t)\xt d t + g(t) d B_t, \qquad t: 0 \to 1, 
\end{align}
where $f(t)$ and $g(t)$ are the drift and diffusion coefficients, and $B_t$ is the standard Brownian motion. 

To generate samples from $\pi(x)$, we simulate a reverse SDE initialized from $x_1 \sim \pi_1(x)$, 
\begin{align}\label{eq:reverse-sde}
d \xt &= \left[f(t) \xt - g(t)^2 \nabla_{x_t} \log \pi_t(x_t) \right] d t + g(t) d \bar B_t, \qquad t:  1 \to 0, 
\end{align}
where $\nabla_{x_t} \log \pi_t(x_t)$ is the score function of the marginal density $\pi_t (x_t)$ at time $t$ induced by the forward process in \Cref{eq:forward-sde}, 
and $\bar B_t$ is the reverse-time Brownian motion.

\paragraph{Diffusion MC samplers.} 
Building on the diffusion model paradigm, recent works use MC methods to simulate the reverse dynamics in \Cref{eq:reverse-sde} whose terminal distribution corresponds to the target distribution $\pi\propto \tilde \pi $ of interest \citep{huang2023reverse,grenioux2024stochastic,he2024zeroth}. While the score function   $\nabla_{x_t} \log \pi_t(x_t)$  is generally intractable, the key idea is to construct MC estimates via the denoising score identity \citep[DSI,][]{vincent2011connection,efron2011tweedie},  
\begin{align}\label{eq:dsi}
\nabla_{x_t}\log \pi_t (x_t) &= \int \frac{\alpha(t) 
\xzero - x_t }{\sigma(t)^2} \pi(\xzero \mid \xt) d \xzero,
\end{align}
where $\pi(\xzero \mid x_t) \propto \tilde \pi(\xzero) \mathcal{N}(\xzero \mid \alpha(t) \xzero, \sigma(t)^2\mathbb{I})$ is the \textit{denoising posterior}, which treats  the unnormalized target $\tilde \pi(\xzero)$ as the prior and the forward transition density $\mathcal{N}(\xzero \mid \alpha(t) \xzero, \sigma(t)^2\mathbb{I})$ from \Cref{eq:forward-sde} as the likelihood. The coefficients $\alpha(t)$ and $\sigma(t)$ are determined by the drift and diffusion terms $f(t)$ and $g(t)$ (see \Cref{app:subsec:diffusion} for details).  

\Cref{eq:dsi} suggests that score estimation can be cast as a posterior inference problem. In practice, one can draw approximate samples from the denoising posterior to form an MC estimate of the score, which is then substituted into the reverse dynamics to generate samples from $\pi$. Other score identities beyond the DSI can also be leveraged (see \Cref{app:subsec:diffusion-based-sampling}). 

\paragraph{Sequential Monte Carlo.}
SMC \citep{naesseth2019elements,chopin2020introduction} is a particle-based method for sampling from a sequence of distributions defined on variables $\xn{0:T}$, terminating at a final target distribution  of interest. 
We consider a reverse-time formulation, where SMC evolves a weighted collection of $N$ particles $\{x_{t}^{(i)}, w_{t}^{(i)}\}_{i=1}^{N}$  from $t=T$ to $0$,  gradually approximating the final target. 

An SMC sampler requires two key design choices \citep{naesseth2019elements}, a sequence of \textit{intermediate proposals}, $q_T(\xT)$  and $\{q_t(\xt \mid \xn{t+1})\}_{t=0}^{T-1}$, and a sequence of (unnormalized) \textit{intermediate target distributions}, $\{\gamma_t(\xn{t:T})\}_{t=0}^T$, such that 
 the final target $\gamma_0(\xn{0:T})$ recovers the  distribution of interest. 

SMC  initializes $N$ particles $\xT^{(i)} \sim q_T(\xn{T})$ with weights $\wT^{(i)} \gets \gamma_T (\xT^{(i)})/q_T(\xT^{(i)})$ for $i=1,\cdots,N$. 
Then for  each step $t = T{-}1,\dots,0$ and particle $i = 1,\dots,N$, it proceeds as follows:
\begin{enumerate}
    \item  resample ancestor $\xn{t+1}^{(i)} \sim \text{Multinomial}\big(\xn{t+1}^{(1:N)}, \wtp^{(1:N)}\big)$;
\item propagate particle $x_t^{(i)} \sim q_t\big(x_t \mid x_{t+1}^{(i)}\big)$;
\item compute weight $\wt^{(i)} \gets \gamma_t(\xn{t:T}^{(i)}) \big/ \left[ \gamma_{+1}(\xn{t+1:T}^{(i)}) q_t(x_t^{(i)} \mid x_{t+1}^{(i)}) \right]$.
\end{enumerate}

The final set of weighted particles forms a discrete approximation to the final target $\gamma_0$, which is asymptotically exact given infinite particles under  regularity conditions \citep{chopin2020introduction}. However, the efficacy of SMC greatly depends on the choice of intermediate targets and proposals -- the closer these intermediate distributions are to the marginals of the final target, the more effective the  SMC sampler.

%% file: sections/method.tex
\section{Method}
Our goal is to sample from a target distribution $\pi (x) =\frac{1}{Z} \tilde \pi(x)$, where $\tilde \pi(x)$ is the unnormalized target and $Z=\int \tilde \pi(x) \mathrm{d}x$ is a generally intractable normalization constant. We develop \ourmethodfull{} (\ourmethod), a diffusion-based SMC sampler targeting $\pi(x)$.  

To enable sequential inference with SMC, we define an \textit{extended target distribution} over a discretized trajectory $\xn{0:T}=(x_0,\dots,x_T)$ of the forward process in \Cref{eq:forward-sde}, 
\begin{align}\label{eq:target_extended}
    \pi(\xn{0:T}) = \pi(x_0) \prod_{t=1}^T \pi(x_t|x_{t-1}),
\end{align}
where $0 =\tau_0< \cdots < \tau_T=1$ are $T+1$ discretization times and 
$\pi(\xt|\xtm)$ is the forward transition density from time $\tau_{t-1}$ to $\tau_t$ induced by \Cref{eq:forward-sde} (see \Cref{app:subsec:diffusion} for analytical expressions). Without loss of generality, we assume a uniform discretization step size $\delta=1/T= \tau_t -\tau_{t-1}, \forall t=1,\cdots, T$.

This construction yields a sequence of intermediate targets $\{\pi(x_{t:T})\}_{t=T}^0$,  whose final marginal $\pi(x_0)$ recovers the desired target. 
These intermediate targets provide a natural setting for SMC, and in fact, are \textit{optimal} intermediate gargets. However, they are intractable to sample from and to evaluate. 

To address this challenge, we develop RDSMC leveraging the dual structure of diffusion processes, generating  samples in the reverse direction while grounding their targets in the forward direction. In  \Cref{subsec:proposal}, we introduce  a sequence of proposals based on the reverse diffusion process using MC score estimates. To correct the resulting proposal bias, in \Cref{subsec:target}, we develop practical intermediate targets that form \textit{exact approximations} to the optimal targets  $\{\pi(x_{t:T})\}_{t=0}^T$ using byproducts of MC score estimates. Finally in \Cref{subsec:alg}, we present the full \ourmethod{} algorithm and its theoretical guarantees.

\paragraph{Notation.} We slightly abuse notation by letting $\xt$ denote both the continuous-time variable $x_t$ for $t \in [0,1]$, as used in \Cref{eq:forward-sde,eq:reverse-sde}, and the discrete-time variable for $t \in \{0,\dots,T\}$, as used throughout this section.  We denote $ \ft\coloneqq f(\tau_t),  \gt \coloneqq g(\tau_t), \alpha_t \coloneqq \alpha(\tau_t), \sigma_t\coloneqq \sigma(\tau_t)$ for time-dependent diffusion coefficients from \Cref{eq:forward-sde,eq:dsi}.  For generality, we define $\ut$ as the collection of  \textit{all} auxiliary random variables generated in the MC score estimation at step $t$.

\subsection{Reverse diffusion   proposal}\label{subsec:proposal}
We design an extended proposal distribution based on the reverse diffusion process with MC score estimates,  jointly modeling diffusion dynamics $\xt$ and  auxiliary randomness $\ut$ from score estimation.

\paragraph{MC score estimation.} Given a sample $\xt$ at step $t$, we define a generic MC score estimator $s(\xt, \ut)\approx \nabla_{\xt} \log \pi(\xt)$  based on the DSI in \Cref{eq:dsi}, where $\ut$ is the estimation randomness associated with some sampling distribution $ q(\ut \mid \xt)$. 

As an illustrative example, consider an importance sampling (IS) estimator (\Cref{alg:importance-sampler}) targeting the posterior $\pi(\xzero \mid \xt) \propto \tilde \pi(\xzero) \mathcal{N}(\xt \mid \alpha_t \xzero, \sigma_t^2 \mathbb{I})$. We draw $M$ importance samples $\utsupm$ from  the proposal $ q(\utsupm \mid \xt) \coloneqq  \mathcal{N}(\utsupm \mid \xt / \alpha_t, \sigma_t^2 / \alpha_t^2 \mathbb{I})$ for $m=1,\cdots, M$, and estimate the score by 
\begin{align}\label{eq:score-is-estimate}
 \nabla_{\xt}  \log \pi(\xt) & \approx   s(\xt, \ut) \coloneqq \sum_{m=1}^M \frac{w^{(m)}}{\sum_{m'=1}^M w^{(m')}} \cdot \frac{\alpha_t \utsupm - x_t}{\sigma_t^2},
\end{align}
which corresponds to an MC approximation of the RHS of \Cref{eq:dsi}. The importance weights $w^{(m)}$ are 
\begin{align}\label{eq:score-is-weight}
w^{(m)} = \frac{\tilde \pi(\utsupm )\, \mathcal{N}(x_t \mid \alpha_t \utsupm, \sigma_t^2 \mathbb{I})}{q(\utsupm \mid \xt)}, \qquad m=1,\cdots, M. 
\end{align}
In this case, the auxiliary randomness is to the collection of all importance samples $\ut=\{\utsupm\}_{m=1}^M$.

In practice, we use a more sophisticated AIS scheme  to improve the accuracy of score estimates (\Cref{alg:annealed-importance-sampler}). More informative IS proposals or other MC methods such as SMC and rejection sampling may also be used; see \Cref{app:subsec:sampling} for further details.

\paragraph{Approximate reverse diffusion dynamics.}
With the score estimates in place, we approximate the  transition kernel of the reverse SDE in \Cref{eq:reverse-sde} to define a conditional proposal for $\xt$, 
\begin{align}\label{eq:qt}
    q(\xt \mid \xtp, \utp) &\coloneqq \mathcal{N}(\xt \mid \xtp - \left[ \ftp \xtp - \gtp^2   s(\xtp, \utp)\right] \delta , \gtp^2 \delta). 
\end{align}

\paragraph{Extended proposal distributions.} 
The full sampling process starts by drawing $\xT$ from  a tractable base distribution $q(\xT)$, e.g. $q(\xT) = \mathcal{N}(0,1)$.  We then iteratively evolve $\xt$ for $t=T-1, \cdots,0$ using the reverse diffusion dynamics $q(\xt \mid \xtp, \utp)$ in \Cref{eq:qt} where auxiliary variable $\utp \sim q(\utp \mid \xtp)$ is sampled in previous iteration to estimate the score $s(\xtp, \utp)$. 

The overall sampling process defines a proposal over the extended space of  $\{\xt, \ut\}_{t=0}^T$, 
\begin{align}\label{eq:proposal}
    q(\xn{0:T},\un{0:T}) &\coloneqq q(x_T) q(\uT \mid \xT) \prod_{t=0}^{T-1} q(x_t \mid x_{t+1}, \utp) q(\ut \mid \xt). 
\end{align}
While sampling $\un{0}$ at the final step $t=0$ is not necessary, we retain it for notation simplicity.

This sampling procedure builds on prior work of  \citet{huang2023reverse,grenioux2024stochastic}, which estimate the score function using MCMC, and by \citet{he2024zeroth} which use rejection sampling. 
We formalize these ideas by defining an extended proposal distribution that incorporates randomness in  MC score estimates. In particular, we  adopt an IS or AIS-based MC approach that also provides an unbiased estimate of the normalization constant, a property that, in principle, can be achieved by the rejection sampling approach of \citet{he2024zeroth} as well. This property will be leveraged in the next part to construct intermediate targets for \ourmethod.

\subsection{Intermediate targets}\label{subsec:target}
Samples from the reverse diffusion proposal deviate from the desired target $\pi$ due to two sources of error: time discretization of the reverse SDE, and bias in score estimation. 
To correct these errors, we design a series of intermediate target distributions. We first characterize the optimal but intractable targets, and then develop practical approximations using byproducts of score estimation. Finally, we extend the intermediate targets to include auxiliary variables. 
Our construction aligns with the optimal targets at intermediate steps and  
recovers the desired marginal $\pi(\xzero)$ at the final step $t=0$.  

\paragraph{Optimal intermediate targets.} 

The optimal intermediate target at step $t$ is the marginal distribution of the extended target $\pi(x_{0:T})$ in \Cref{eq:target_extended}, 
\begin{align}\label{eq:optimal-target}
\pi(\xn{t:T}) = \int \pi(\xn{0:T}) \mathrm{d} \xn{0:t-1} = \pi(\xt) \prod_{i=t+1}^T \pi (\xn{i} \mid \xn{i-1}), 
\end{align}
as it leads to exact samples from $\pi(x_{0:T})$ when combined with locally optimal proposals \citep{naesseth2019elements}. 

However, the marginal $\pi(x_t)=\int \pi(\xzero) \mathcal{N}(\xt \mid \alpha_t \xzero, \sigma_t^2 \mathbb{I}) \mathrm{d} \xzero$ in \Cref{eq:optimal-target}  involves an intractable integral, except in the final step $t=0$, where \(\pi(x_0)\) is known up to a normalization constant.

\paragraph{Marginal estimation.}
For $t>0$, we approximate the marginal  $\pi(\xt)$ by $\hat \pi(\xt, \ut)$ using 
byproducts of MC score estimates, where we recall $\ut$ is the estimation  randomness. The key insight is that $Z \pi(\xt)$ is the normalization constant of the posterior  $\pi(\xzero \mid \xt)\propto \tilde \pi(\xzero) \mathcal{N}(\xt \mid \alpha_t \xzero, \sigma_t^2 \mathbb{I})$, that  is, 
\begin{align*}
    \int \tilde \pi(\xzero) \mathcal{N}(\xt \mid \alpha_t \xzero, \sigma_t^2 \mathbb{I}) \mathrm{d} \xzero &= Z \int \pi(\xzero) \mathcal{N}(\xt \mid \alpha_t \xzero, \sigma_t^2 \mathbb{I}) \mathrm{d} \xzero  = Z \pi(\xt). 
\end{align*}
Hence, we can re-use the posterior inference procedure   in  MC score estimation to obtain an unbiased estimate of the desired marginal $\pi(\xt)$ (up to a factor of $Z$). 

For example, IS or AIS-based score estimation (\Cref{alg:importance-sampler,alg:annealed-importance-sampler}) generates importance weights \(\{w^{(m)}\}_{m=1}^M\)  targeting $\pi(\xzero \mid \xt)$ for a fixed $\xt$; we then obtain an unbiased marginal estimate 
\begin{align}\label{eq:marginal-estimate}
\hat \pi(\xt, \ut) &\coloneqq \frac{1}{M} \sum_{m=1}^M  w^{(m)},  \textrm{ with } \mathbb{E}_{q(\ut \mid \xt)}\left[ \hat \pi (\xt, \ut) \right] =Z \pi(\xt). 
\end{align}

At the final step $t=0$, we set $\hat \pi (\xzero, \uzero)\coloneqq \tilde \pi(\xzero) = Z \pi(\xzero)$  as no approximation is needed.

\paragraph{Extended intermediate targets.}
Finally, we incorporate the auxiliary variables $\ut$ by setting their targets to match their sampling distributions and  
 define the extended intermediate target as 
\begin{align}\label{eq:intermediate-target}
    \gamma_t(\xn{t:T}, \un{t:T}) &\coloneqq
\hat \pi(\xt, \ut) \, q(\ut \mid \xt) \prod_{i=t}^{T-1} \pi (\xn{i+1} \mid \xn{i}) \, q(\un{i+1} \mid \xn{i+1})
\end{align}
for $t=0,\cdots, T-1$, and $\gamma_T(\xT, \uT) \coloneqq  \hat \pi (\xT, \uT) \, q(\uT \mid \xT)$. 

The structure in \Cref{eq:intermediate-target} mirrors that of the optimal targets in \Cref{eq:optimal-target}, replacing the intractable marginal $\pi (\xt)$ with an unbiased estimate $\hat \pi (\xt, \ut)$ (up to a factor of $Z$) and accounting for auxiliary randomness $\un{t:T}$. 
Moreover, we make the following two observations. 

\begin{observation}\label{obs:final-target} 
 The final marginal target at $t=0$  matches the desired $\pi(\xzero)$ with \emph{no approximations}, 
\begin{align}\label{eq:final-marginal}
\begin{split}
\gamma_0(\xzero)
&= \int \gamma_0(\xn{0:T}, \un{0:T})  \mathrm{d} \xn{1:T}, \un{0:T} \\
&=\int \hat \pi( \xzero, \uzero)    q(\uzero \mid \xzero) \prod_{i=1}^T \pi (\xn{i} \mid \xn{i-1}) \, q(\un{i} \mid \xn{i})  \mathrm{d} \xn{1:T}, \un{0:T}  \\
& = \tilde \pi (\xzero) \propto \pi(\xzero), 
\end{split}
\end{align}
as  $\hat \pi (\xzero, \uzero)=\tilde \pi (\xzero)$ by construction. This implies our final SMC target is correctly specified. 
\end{observation}

\begin{observation}\label{obs:exact-approximation}
Marginalizing out $\un{t:T}$ in $\gamma_t(\xn{t:T}, \un{t:T})$ recovers the optimal target $\pi(\xn{t:T}), \forall t$,   
\begin{align}\label{eq:intermediate-marginal}
\begin{split}
    \int  \gamma_t(\xn{t:T}, \un{t:T}) \mathrm{d} \un{t:T}  &= \int \hat \pi(\xt, \ut) \, q(\ut \mid \xt) \prod_{i=t}^{T-1} \pi (\xn{i+1} \mid \xn{i}) \, q(\un{i+1} \mid \xn{i+1}) \mathrm{d} \un{t:T } 
    \\ 
 &=  \mathbb{E}_{q(\ut \mid \xt)}\left[ \hat \pi(\xt, \ut) \right] \, \prod_{i=t}^{T-1} \pi(\xn{i+1} \mid \xn{i})  \\
    &=   Z \pi(\xt) \, \prod_{i=t}^{T-1} \pi(\xn{i+1} \mid \xn{i})  \propto \pi (\xn{t:T}), 
\end{split}
\end{align}
where  the last equality follows from the unbiasedness of  marginal estimates, as in \Cref{eq:marginal-estimate}. 
This result shows that our intermediate targets  $\gamma_t(\xn{t:T}, \un{t:T})$ are aligned with the optimal $\pi(\xn{t:T})$, despite the  approximation in $\hat \pi(\xt, \ut)$, a property known as the \emph{exact approximation} \citep{andrieu2010particle,naesseth2019elements}. 

\end{observation}

While the generic SMC framework remains valid under alternative choices of intermediate targets, provided that  $\gamma_0(\xzero)\propto \pi(\xzero)$, our construction offers a balance between theoretical correctness and practical efficiency. The  marginal estimates $\hat \pi(\xt, \ut)$ in our intermediate targets from \Cref{eq:intermediate-target}  act as  \textit{twisting} or \textit{look-ahead} functions \citep[][Chapter 3]{naesseth2019elements} that incorporate future information $\pi(\xzero)$ into intermediate steps. 
Together, \Cref{obs:final-target,obs:exact-approximation} ensure that our intermediate  targets provide effective guidance throughout the sampling process while recovering the desired target in the end.

\subsection{\ourmethod: Algorithm and theoretical guarantee}\label{subsec:alg}
We now derive weighting functions, the final component of \ourmethod{}, which guides the resampling of particles. 
We then present the complete algorithm and establish its theoretical guarantees. 

\input{algorithms/ourmethod}

\paragraph{Weighting functions.} Following the standard SMC framework, we  define the weighting functions  on the extended space $\{\xn{t:T}, \un{t:T}\}$ for $t=T-1,\cdots, 0$ as ratios of  intermediate targets and proposals 
\begin{align}\label{eq:weight}
    w_t & \coloneqq \frac{\gamma_t(x_{t:T}, \un{t:T})}{\gamma_{t+1}(\xn{t+1:T}, \un{t+1:T})\,q(\xt  \mid \xtp, \utp) \, q(\ut \mid \xt)}. 
\end{align}
Substituting the expressions for $\gamma_t$ from \Cref{eq:intermediate-target} and canceling out common terms yields   
\begin{align}\label{eq:weight-final}
     w_t  & = \frac{\hat \pi (\xt,\ut) \pi(\xtp|\xt)}{\hat \pi (\xtp,\utp) q(\xt  \mid \xtp, \utp) }.
\end{align}

At the initial step \(T\), the weight is 
$w_T \coloneqq \frac{\gamma_T(\xT, \uT)}{q (\xT) q(\uT \mid \xT)} = \frac{\hat \pi(\xT, \uT)}{q(x_T)}$ recalling that $\gamma_T(\xT, \uT)= \hat \pi(\xT, \uT) q(\uT \mid \xT)$ by construction. 

Notably, although the intermediate targets and proposals involve auxiliary sampling distributions $q(\ut \mid \xt)$ for score estimation, the weighting functions $w_t$ in \Cref{eq:weight-final} and $w_T$ do \textit{not} depend on their evaluation. Hence, a generic MC sampler can be used while still retaining  computable weights, as long as it produces tractable estimates of the score and marginal densities.

\paragraph{The \ourmethod{} algorithm and theoretical guarantees.}  
We summarize the \ourmethod{} algorithm in \Cref{alg:main}. 
It initializes $N$ particles from a tractable base distribution. Subsequently, the particles are propagated through a reverse diffusion-based proposal, followed by weighting and resampling using intermediate targets. These steps are enabled by running an inner-level MC estimation of the score function and the marginal density. In this work, we use AIS (\Cref{alg:annealed-importance-sampler}) while  other methods can be incorporated as well  (see \Cref{app:subsec:sampling}). The final output of \ourmethod{} is a weighted set of samples approximating the target $\pi(\xzero)$, along with an estimate of the normalization constant $Z$. 

\ourmethod{} can be viewed as an adaptation of the \textit{nested} SMC framework \citep{naesseth2015nested,naesseth2019high}. The inner-level MC estimation involves a sampling procedure (e.g. AIS) targeting the intractable posterior $\pi (\xzero \mid \xt)$, 
which is then used to assign \emph{proper weights} \citep[Chapter 4.3]{naesseth2019elements} to proposed samples $x_t$. 
Consequently,  \ourmethod{}  inherits the unbiasedness and asymptotic exactness guarantees  of nested SMC \citep{naesseth2015nested,naesseth2019elements}. 

\begin{theorem}[Informal]\label{thm:informal} Under regularity conditions,  
    the \ourmethod{} algorithm provides a consistent estimator of the target distribution $\pi(x_0)$ as particle size  $N\to\infty$ and an unbiased estimator of the normalization constant $Z$ for any $N\geq 1$.
\end{theorem}

We provide the formal statement and proof in \Cref{app:sec:theory}.

\Cref{thm:informal} shows that \ourmethod{} effectively wraps a diffusion MC-style proposal within an SMC framework, mitigating its bias as the number of particles increases. Moreover, it provides an unbiased estimate of the normalization constant, a property absent in existing diffusion MC methods \citep{huang2023reverse,grenioux2024stochastic,he2024zeroth}. 

The bias correction mechanism of \ourmethod{} arises from two key aspects: (1) the final target of \ourmethod{} is explicitly constructed to match the desired target $\pi(\xzero)$, regardless of the discretization scheme or marginal density approximation (\Cref{obs:final-target}); and (2) errors in score estimation affect only the proposal steps, which are accounted for in the weighting functions. Therefore, the weighting and resampling steps in the outer SMC loop asymptotically correct for proposal bias, ensuring convergence to the final target $\pi(\xzero)$ in the limit of many particles.

%% file: algorithms/ourmethod.tex
\begin{algorithm}[!t]
\caption{\ourmethodfull{} (\ourmethod)}
\label{alg:main}
\KwIn{Unnormalized target $\tilde \pi(\xzero)$, number of particles $N$,  discretization steps $T$ (with step size $\delta=1/T$),   diffusion schedule $\{\alpha_t, \sigma_t, \ft, \gt\}$, base distribution $q(\xT)$, and additional inputs for score and marginal estimation $\eta$} 
\KwOut{Weighted samples $\{x_0^{(i)}, w_0^{(i)}\}_{i=1}^N$ and normalization constant estimate $\hat Z$}

\For{$i \gets 1$ \KwTo $N$}{
    Sample $\xT^{(i)} \sim q(\xT)$\;
    Compute score and marginal estimates $s_T^{(i)}, \hat \pi_T^{(i)} \gets$  Algorithm 3$(\pi, \alpha_T, \sigma_T, \xT^{(i)}, \eta)$\;
     Compute weight: $\wT^{(i)} \gets \frac{\hat \pi_T^{(i)}}{ q (\xT^{(i)})}$\;
}

\For{$t \gets T-1$ \KwTo $0$}{
    Resample $\{\xtp^{(i)}, \utp^{(i)}, s_{t+1}^{(i)}, \hat \pi_{t+1}^{(i)}\}_{i=1}^N$ according to weights $\{\wtp^{(i)}\}_{i=1}^N$ \;
    \For{$i \gets 1$ \KwTo $N$}{
         Sample \small{$x_t^{(i)} \sim q(\xt \mid \xtp^{(i)}, \utp^{(i)}):=\mathcal{N}\left(\xt \mid \xtp^{(i)} - \left[ \ftp \xtp^{(i)} - g_{t+1}^2 s_{t+1}^{(i)}\right] \delta , \gtp^2 \delta \right)$}\;
         \If{t > 0}{
        Compute score and marginal estimates $s_t^{(i)}, \hat \pi_t^{(i)} \gets $ Algorithm 3 $(\pi, \alpha_t, \sigma_t, \xt^{(i)}, \eta)$\;}
        \Else{
        Compute exact marginal $\hat \pi_0^{(i)} \gets \tilde \pi (\xzero^{(i)})$\;
        }
        Compute weight: $\wt^{(i)} \gets \frac{\hat \pi_t^{(i)}\,  \pi(\xtp^{(i)} \mid \xt^{(i)})}{\hat \pi_{t+1}^{(i)} \,  q(\xt^{(i)} \mid \xtp^{(i)}, \utp^{(i)})}$\; \label{algline:weight}
    }
}
Compute  normalization constant estimate $\hat Z \gets \prod_{t=0}^T \frac{1}{N}\sum_{i=1}^N \wt^{(i)}$.
\end{algorithm}

%% file: sections/related_works.tex
\section{Related works}\label{sec:related-works}

\paragraph{Diffusion MC samplers.}
\citet{huang2023reverse} develop the RDMC sampler based on the reverse diffusion process using Langevin-style MCMC to estimate the score function. They initialize the sampler at some $\tau_T < 1$  via a nested Langevin procedure. \citet{grenioux2024stochastic} propose a similar SLIPS sampler with a signal-to-noise-ratio (SNR)-adjusted discretization scheme, and provide further guidance on choosing the initial sampling time $\tau_T$ under suitable conditions. In contrast, we use AIS-based score estimators to enable marginal estimation and start sampling from the base distribution at $\tau_T=1$.  

Alternatively, \citet{he2024zeroth} explore rejection sampling for score estimation, improving the  sampling efficiency in low-dimensional settings, which may be incorporated into our framework as well. 

While these methods provide certain theoretical guarantees,  we offer an orthogonal means to improve their accuracy by increasing the number of particles. \ourmethod{} can be viewed as a generic SMC wrapper around these diffusion-MC samplers (given suitable score estimators),  enabling  consistent sampling and unbiased estimation of the normalization constant.

\paragraph{Training-based diffusion samplers} 
In addition to developing MC estimates of the score function, another line of work trains a neural network directly using unnormalized target information. Variational approaches achieve this goal by minimizing the divergence between forward and reverse processes \citep{zhang2021path,berner2022optimal,vargas2023denoising,vargas2023transport,richter2023improved}, while  \citet{akhound2024iterated,ouyang2024bnem} exploit various identities of the score function or a related energy function. Similar to their training-free MC-based counterparts, these methods are prone to discretization and score approximation errors.

Closest to our work is \citet{phillips2024particle} that also use SMC to correct proposal bias. Their method requires training neural networks,  while ours is training-free. Moreover, their neural network-based intermediate targets incur approximation errors except at the last step, while  ours reflect the true target marginals at each SMC iteration. See an empirical comparison in \Cref{app:subsubsec:ablation:pdds}. 

\citet{he2025no} show that many training-based methods incur heavy computational overhead compared to classical sampling methods and rely on special network preconditioning. For this reason, we limit our empirical comparison to training-free samplers.  Nonetheless, an interesting direction for future work is to combine our method with training-based approaches to further enhance performance.

\paragraph{SMC for conditional generation from diffusion models.} 
Beyond classical sampling tasks, SMC has been applied to conditional generation for pre-trained diffusion models \citep{trippe2022diffusion,wu2023practical,cardoso2023monte,singhal2025general}. 
While these works also combine SMC and diffusion dynamics, a key distinction is the target distribution:  they sample from an existing diffusion model prior tilted with a reward function, whereas our setting involves sampling from any (unnormalized) target distribution. As a result, the way diffusion models are used is different, and we require distinct designs for  proposals and intermediate targets in SMC. 

Nevertheless, \ourmethod{}  uses the estimated marginal densities in \Cref{eq:intermediate-target} as twisting functions  to improve SMC sampling efficiency. This strategy is conceptually is similar to that of \citet{wu2023practical}, though in a different setting;  see \Cref{app:sec:related-works} for further discussion. 

%% file: sections/experiments.tex
\section{Experiments}
We evaluate \ourmethod{} on a range of  synthetic and real-world target distributions, comparing it to SMC \citep{del2006sequential}, AIS \citep{neal2001annealed}, SMS \citep{saremi2023chain}, RDMC \citep{huang2023reverse}, and SLIPS \citep{grenioux2024stochastic}. AIS and SMC operate over a series of geometric interpolations between the target and a Gaussian proposal with MCMC transitions. 
SMS samples from the target by sequentially denoising 
equally noised measurements using a  Langevin procedure combined with estimated scores.  RDMC and SLIPS are described in \Cref{sec:related-works}.

\ourmethod{} uses a variance-preserving diffusion schedule. To evaluate the effectiveness of our intermediate targets and the importance of resampling, we include two ablations: (i) \ourmethod{} (Proposal), which  samples directly from the reverse diffusion proposal in \Cref{eq:proposal} without any weighting or resampling, and (ii) \ourmethod{} (IS), which applies a final IS correction to samples from \ourmethod{} (Proposal). 

Baseline methods follow the implementation of \citet{grenioux2024stochastic}. 
Notably, the information about the target variance (or an estimation) is provided to guide the initialization of the baselines SMC, AIS, SMS and SLIPS. In contrast, our method does \textit{not} make use of this extra information.

Unless otherwise specified, we use $T=100$ discretization steps for \ourmethod{} and its variants, and $T=1,024$ steps for other methods. We generate $N=4,096$ final samples for all methods, and tune their hyperparameters assuming access to either a validation dataset or an oracle metric. 

We provide ablation studies controlling for the discretization steps, the total running time and comparable hyperparameter settings in \Cref{app:subsec:ablation}, where we observe \ourmethod{} still remains competitive and, in some cases, superior. All experiment details are included in \Cref{app:sec:experiments}. 

\subsection{Bi-modal gaussian mixtures}

\begin{figure}[!t]
  \centering

  \begin{subfigure}[t]{0.28\textwidth}
    \centering
    \includegraphics[width=\linewidth]{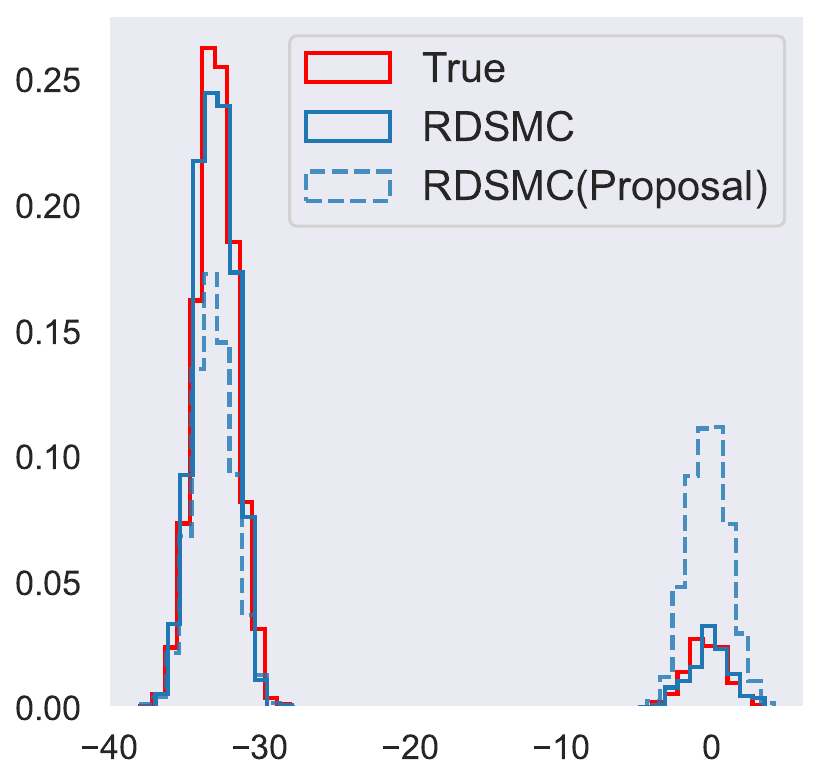}
    \caption{$d=2$:  histogram along 1st dimension. While \ourmethod{}'s proposal covers the modes, the SMC procedure is crucial for producing calibrated weights.}
    \label{subfig:gmm-hist}
  \end{subfigure}
  \hfill
  \begin{subfigure}[t]{0.7\textwidth}
    \centering
      \includegraphics[width=\linewidth]{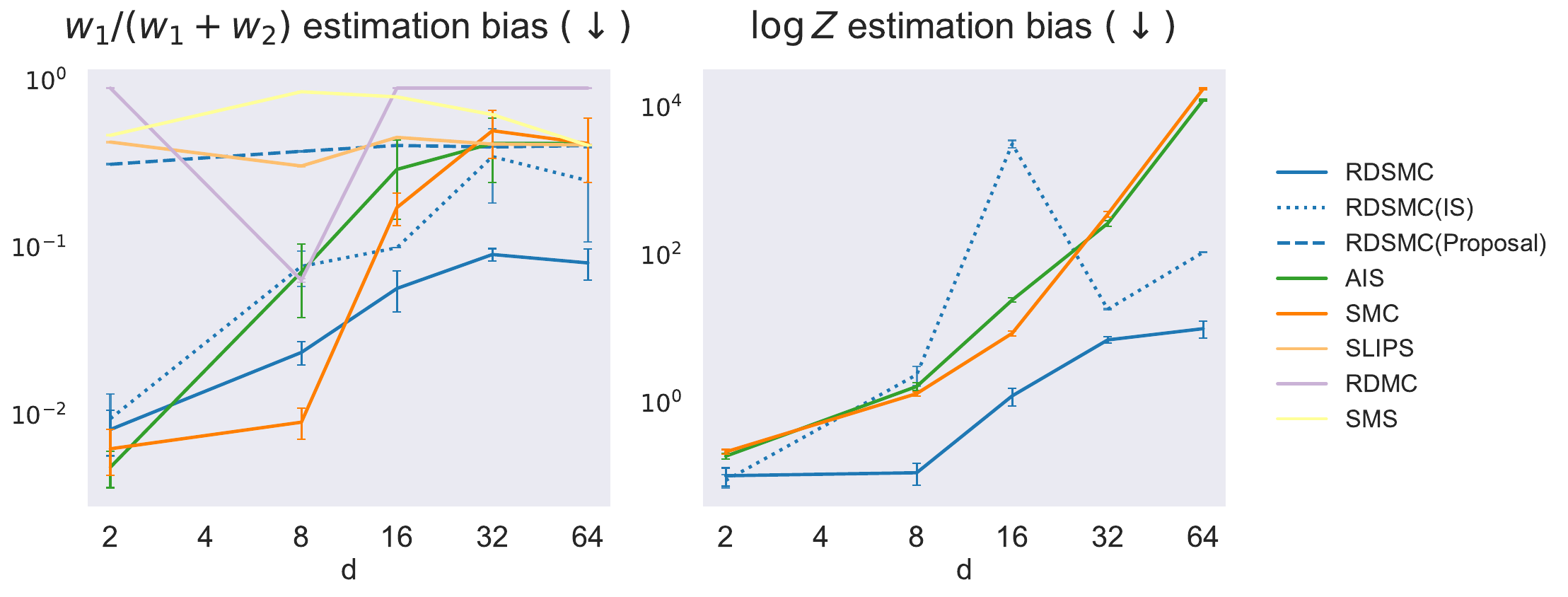}
      \caption{Estimation bias of weight ratio $w_1/(w_1+w_2)$ and log-normalization constant $\log Z$ versus dimension $d$. Results are averaged over 5 seeds with error bars showing one standard error. Note that SMS, RDMC, and SLIPS do not provide estimate of $\log Z$. For both metrics, the bias increases with $d$ for all methods, while \ourmethod{} outperforms the baselines for most cases. }
      \label{subfig:gmm-bias-vs-d}
  \end{subfigure}
  \caption{Bi-modal Gaussian mixture study.}
  \label{fig:gmm}
\end{figure}

We first study bi-modal Gaussian mixtures with an imbalanced weight ratio of $w_1 / (w_1+w_2) = 0.1$ for varying dimensions $d$. The estimated ratio is obtained by assigning samples to the mode with the highest posterior probability. 
For each method we select the hyperparameters based on the lowest estimation bias of the weight ratio. 

In \Cref{subfig:gmm-hist}, we compare the marginal histogram of samples from \ourmethod{}, \ourmethod{} (Proposal), and the true target in $d=2$.  While samples from \ourmethod{} (Proposal) cover both modes, their relative weights are overly balanced. In contrast, \ourmethod{} recovers the calibrated mode weights, indicating the importance of the error correction mechanism by SMC.

\Cref{subfig:gmm-bias-vs-d} shows the estimation bias of the weight ratio  (left), and that of the log-normalization constant $\log Z$ (right).  Note that only \ourmethod{}, \ourmethod{}(IS), AIS and SMC  provide estimates for $\log Z$. We observe that \ourmethod{} consistently outperforms other methods in both metrics across dimensions. Moreover, \ourmethod{} (Proposal) exhibits consistently high  weight ratio estimation bias. While \ourmethod{} (IS) reduces this bias, it still underperforms the full \ourmethod{} procedure, highlighting the effectiveness of intermediate resampling guided by our intermediate target distributions. 

In high dimensions, all methods exhibit some degree of mode collapse. \ourmethod{} primarily samples from the dominant mode, resulting in average weight ratio estimation biases of  0.09 and 0.08 for $d=32$ and $64$, respectively. In contrast, AIS and SMC completely collapse to a single mode, which varies across runs, yielding degenerate estimated ratios of 0 or 1 and an average bias of around 0.4.

\subsection{Rings and Funnel distributions}
\begin{table}[!t]
    \centering
    \input{results/tab_rings_and_funnel}
    \vspace{0.5em}
    \caption{Results on Rings and Funnel (mean $\pm$ standard error over 5 seeds). Bold indicates 95\% confidence interval overlap with the best average result. \ourmethod{} has the lowest $\log Z$ estimation bias for both targets. On Rings, \ourmethod{}(Proposal) has the lowest radius TVD,  followed by AIS, SMC, and \ourmethod{}. On Funnel, SLIPS has the lowest sliced KSD, followed by AIS, SMC, and \ourmethod{}.}
    \label{tab:rings-and-funnel}
\end{table}

We present results on two additional synthetic targets. 
Rings, introduced by \citet{grenioux2024stochastic}, is a 2-dimensional distribution constructed via an inverse polar parameterization: the angular component is uniformly distributed, while the radius component follows a 4-mode Gaussian mixture. Funnel \citep{neal2003slice}, is a 10-dimensional  ``funnel"-shaped distribution.

For Rings, we assess sample quality using  total variational distance in the radius component (Radius TVD). For Funnel, use sliced Kolmogorov-Smirnov distance (Sliced KSD) to evaluate sample quality following \citet{grenioux2024stochastic}.  
As both targets admit tractable normalization constants,  we also report the estimation bias  for methods that compute them. For each method, we select the hyperparameters with lowest Raidus TVD on a heldout validation set for Rings and lowest Sliced KSD for Funnel. 

As shown in \Cref{tab:rings-and-funnel}, \ourmethod{} achieves lower or comparable estimation bias of the log-normalization constant relative to AIS and SMC on both targets. 
On the lower-dimensional Rings target, \ourmethod{} (Proposal) has the lowest radius TVD, where AIS, SMC and \ourmethod{} closely match. On the more challenging Funnel target, SLIPS achieves the lowest sliced KSD; while \ourmethod{} performs slightly worse, it greatly outperforms its IS and Proposal variants.

\subsection{Bayesian logistic regression}

\setlength{\tabcolsep}{3pt}
\begin{table}[!t]
    \centering
\input{results/tab_logistic_regression}

\vspace{1em}
    \caption{Bayesian logistic regression with test log-likelihood (mean $\pm$ standard error) averaged over 5  seeds. Bold indicates 95\% confidence interval overlap with that of the best average result. SLIPS achieves the best overall performance, while \ourmethod{} matches or closely approaches it. In contrast, \ourmethod{}(Proposal) performs noticeably worse, highlighting the effectiveness of the SMC correction.} 
    \label{tab:logistic-regression}
\end{table}
\setlength{\tabcolsep}{6pt}

Finally, we evaluate inference performance on  Bayesian logistic regression models using four datasets. Credit and Cancer \citep{nishihara2014parallel} involve predicting credit risk and breast cancer recurrence, while Ionosphere \citep{sigillito1989classification} and Sonar \citep{gorman1988analysis} focus on classifying radar and sonar signals, respectively. 
The inference is performed on $60\%$ of each dataset, leaving  
20\% for validation and 20\% for testing. Each method is tuned using the validation log-likelihood estimate. We report the test log-likelihood in \Cref{tab:logistic-regression}. 

\ourmethod{} achieves the best or near-best performance across datasets, closely matching SLIPS, which performs the best overall. This comparison suggests that SLIPS's time discretization and initialization strategy may offer complementary benefits despite lacking systematic error correction. \ourmethod{} (IS) underperforms on Credit and Cancer and exhibits higher variance than \ourmethod{}, while \ourmethod{} (Proposal) performs the worst, again highlighting the importance of the SMC mechanism. Compared to baseline methods, \ourmethod{} shows greater variance, likely due to auxiliary randomness in its nested structure, suggesting a direction for future improvement.

%% file: results/tab_rings_and_funnel.tex
\begin{tabular}{lcccc}
\toprule
\multirow{2}{*}{Algorithm} & \multicolumn{2}{c}{Rings ($d=2$)} & \multicolumn{2}{c}{Funnel ($d=10$)} \\
\cmidrule(lr){2-3} \cmidrule(lr){4-5}
& Radius TVD  $(\downarrow)$& $\log Z$ Bias  $(\downarrow)$ & Sliced KSD   $(\downarrow)$ & $\log Z$ Bias  $(\downarrow)$ \\
\midrule
AIS & 0.10 $\pm$ 0.00 & 0.05 $\pm$ 0.00 & 0.07 $\pm$ 0.00 & \textbf{0.28 $\pm$ 0.01} \\
SMC & 0.10 $\pm$ 0.00 & 0.05 $\pm$ 0.00 & 0.07 $\pm$ 0.00 & \textbf{0.28 $\pm$ 0.01} \\
SMS & 0.24 $\pm$ 0.01 & N/A & 0.15 $\pm$ 0.00 & N/A \\
RDMC & 0.37 $\pm$ 0.00 & N/A & 0.13 $\pm$ 0.00 & N/A \\
SLIPS & 0.19 $\pm$ 0.00 & N/A & \textbf{0.06 $\pm$ 0.00} & N/A \\
\hline
\ourmethod & 0.13 $\pm$ 0.01 & \textbf{0.03 $\pm$ 0.00} & 0.11 $\pm$ 0.00 & \textbf{0.28 $\pm$ 0.10} \\
\ourmethod(IS) & 0.15 $\pm$ 0.01 & \textbf{0.02 $\pm$ 0.01} & 0.33 $\pm$ 0.03 & 1.61 $\pm$ 0.14 \\
\ourmethod(Proposal) & \textbf{0.09 $\pm$ 0.00} & N/A & 0.32 $\pm$ 0.03 & N/A \\
\bottomrule
\end{tabular}

%% file: results/tab_logistic_regression.tex
 
\begin{tabular}{lrrrr}
\hline
Test LL $(\uparrow)$ & Credit ($d=25$) & Cancer ($d=31$) & Ionosphere ($d=35$) & Sonar ($d=61$) \\
\hline
AIS & $\mathbf{-122.73 \pm 0.51}$ & $-60.45 \pm 0.31$ & $\mathbf{-86.37 \pm 0.10}$ & $-110.11 \pm 0.06$ \\
SMC & $-123.17 \pm 0.05$ & $-60.28 \pm 0.11$ & $\mathbf{-86.37 \pm 0.10}$ & $-110.11 \pm 0.06$ \\
SMS & $-527.79 \pm 0.85$ & $-215.64 \pm 0.66$ & $-202.56 \pm 0.16$ & $-275.44 \pm 0.31$ \\
RDMC & $-388.24 \pm 1.75$ & $-182.81 \pm 0.43$ & $-108.67 \pm 0.09$ & $-128.29 \pm 0.03$ \\
SLIPS & $\mathbf{-121.79 \pm 0.04}$ & $\mathbf{-56.26 \pm 0.08}$ & $\mathbf{-85.07 \pm 0.07}$ & $\mathbf{-102.39 \pm 0.03}$ \\
\hline
\ourmethod & $\mathbf{-124.00 \pm 1.96}$ & $-62.23 \pm 1.96$ & $\mathbf{-87.72 \pm 1.75}$ & $\mathbf{-101.52 \pm 1.84}$ \\
\ourmethod(IS) & $-144.38 \pm 4.63$ & $-82.47 \pm 7.78$ & $\mathbf{-84.90 \pm 2.84}$ & $\mathbf{-110.57 \pm 4.54}$ \\
\ourmethod(Proposal) & $-606.03 \pm 1.26$ & $-246.86 \pm 0.97$ & $-92.62 \pm 0.04$ & $-134.72 \pm 0.13$ \\
\hline
\end{tabular}

%% file: sections/discussion.tex
\section{Discussion}

This paper presents a training-free and diffusion-based SMC sampler, \ourmethodfull{} (\ourmethod), for sampling from unnormalized distributions. 
By leveraging reverse diffusion dynamics as proposals, we devise informative intermediate targets to correct systematic errors within a principled SMC framework. 
These components rely on Monte Carlo score estimation, without requiring additional training. 
Our algorithm provides asymptotically exact samples from the target distribution and a finite-sample unbiased estimate of the normalization constant. 
Empirical results on both synthetic and Bayesian inference tasks demonstrate competitive or superior performance compared to existing diffusion-based and classical geometric annealing-based samplers. 
 
\paragraph{Limitations and future work.} As a nested SMC procedure, \ourmethod{}  introduces auxiliary variance. Such variance may be  mitigated by enhancing autocorrelation within the MC estimation step \citep[e.g.][]{deligiannidis2018correlated}.
In high-dimensional Gaussian mixture experiments, we observe a degree of mode collapse, an issue also seen in other samplers. Potential remedies include  partial resampling  \citep{martino2016weighting}, or using more informed proposals. 
Our method also relies on oracle metrics for hyperparameter tuning, which aligns with existing works; however, developing a more automated strategy remains an important future direction. Finally, 
to further improve performance, we can  
incorporate SNR-adapted discretization schemes and informative initializations, as explored by  \citet{grenioux2024stochastic}, as well as combine our approach with training-based diffusion samplers.

%% file: sections/appendix.tex
\appendix
\input{sections/appendix_background}
\input{sections/appendix_method}

\input{sections/appendix_theory}

\input{sections/appendix_related_works}
\input{sections/appendix_experiments}

%% file: sections/appendix_background.tex
\section{Background: diffusion models and diffusion-based samplers}\label{app:sec:background}

\subsection{Diffusion models}\label{app:subsec:diffusion}
We review common diffusion schedules used in the literature. 
\paragraph{Variance-preserving schedule.}
In Variance-Preserving (VP) schedule \citep{ho2020denoising}, the coefficients of the forward SDE in \Cref{eq:forward-sde} are given by
\begin{align}\label{eq:vp}
f(t) = -\frac{1}{2} b(t), \quad g(t) = \sqrt{b(t)},
\end{align}
where \(b(t)\) is a time-dependent function 
\[
b(t) = b_{\min} + t (b_{\max} - b_{\min}), 
\]
for some constants $b_{\min} \approx 0 \ll b_{\max}$. For a large value of $b_{\max}$, the reference distribution is approximately standard normal, $p_1(x) \approx \mathcal{N}(x  \mid 0, 1)$.

By Ito's calculus (see e.g. \citet{sarkka2019applied}), one can show that the forward transition density $\pi(\xn{s} \mid \xt)$ for $1\geq s>t\geq 0$ is given by 
\[ \pi(\xn{s} \mid \xt) = \mathcal{N}\left(\xn{s} \mid \frac{\alpha(s)}{\alpha(t)} \xt, \left(1-\frac{\alpha^2(s)}{\alpha^2(t)}\right) \mathbb{I} \right),
\]
where 
\[
\alpha(t) = \exp\left(-\frac{1}{2} \int_0^t b(s) ds\right), \quad
\sigma(t)^2 = 1- \exp\left(-\int_0^t b(s) ds\right).
\]

\paragraph{Variance-exploding schedule.} In Variance-Exploding (VE) schedule \citep{song2020score}, the forward SDE coefficients in \Cref{eq:forward-sde} are given by
\[
f(t)=0, \qquad g(t) = \sqrt{ \frac{\mathrm{d} \sigma^2(t)}{\mathrm{d} t}}, 
\]
where $\sigma(t)$ is a non-negative and monotone increasing function with $\sigma(0)=0$. Commonly used forms of $\sigma(t)$ include polynomial functions, such as linear, quadratic or exponential schedules. When $\sigma(1)$ is large, the reference distribution can be effectively approximated as $\pi_1(\xone) \approx \mathcal{N}(\xone \mid 0, \sigma^2(1))$.

Similarly to the VP case, one can derive the following parameters of the forward transition kernel with $\alpha(t)=1$ and $\sigma(t)$ is the given non-negative and monotone function.

\subsection{Score identities used for diffusion MC samplers}
\label{app:subsec:diffusion-based-sampling}

In addition to the DSI from \Cref{eq:dsi}, one may consider the following identities to form alternative MC estimates of the score, given MC samples of $\pi_t(\xzero \mid \xt)$. 

\paragraph{Target score identity (TSI)}
TSI utilizes known structural properties of the target density, such as estimated gradients, facilitating potentially more accurate and lower-variance score estimations at low-noise settings. Formally, the score function of the noisy target distribution $\pi_t(x)$ under the TSI framework satisfies
\begin{align}\label{eq:tsi}
\nabla_{x}\log \pi_t (x) &= \frac{1}{\alpha(t)}\int \nabla_x\log \pi_t(x_0) \pi_t(x_0 \mid x) dx_0.
\end{align}

TSI and its related alternative have been used in several recent studies to improve score estimation accuracy \citep{debortoli2024targetscorematching,  phillips2024particle, akhound2024iterated,he2025featfreeenergyestimators}. 

\paragraph{Mixed score identity (MSI)}

To balance the advantages of DSI and TSI across different noise regimes, MSI combines both identities through a convex interpolation. Specifically, MSI favors TSI in the low-noise regime and DSI in the high-noise regime, leveraging both strengths. Formally, MSI is a convex combination of TSI and DSI with coefficients $\frac{\alpha(t)^2}{\alpha(t)^2 + \sigma(t)^2}$ and $\frac{\sigma(t)^2}{\alpha(t)^2 + \sigma(t)^2}$:
\begin{align}\label{eq:msi}
\nabla_{x}\log \pi_t (x) &= \frac{1}{\alpha(t)^2 + \sigma(t)^2}\int \Big(\alpha(t) \big(x_0 + \nabla_x\log \pi_t(x_0)\big) -x(t) \Big) \pi_t(x_0 \mid x) dx_0.
\end{align}

\citet{he2024training} leverage MSI to train an implicit generative model that produces samples in a single step, demonstrating improved sample quality and efficiency.

%% file: sections/appendix_method.tex
\section{Method details}\label{app:sec:method}
In this section, we discuss several details of \ourmethod{}, including  score and marginal estimation strategies in \Cref{app:subsec:sampling}, the computational complexity in \Cref{app:subsec:method:complexity}, and implementation guidelines in \Cref{app:subsec:method:implement}. 

\subsection{Score  and marginal estimation}\label{app:subsec:sampling}

In this subsection, we present several score and marginal estimators that can be used within \ourmethod{}, which are based on IS, AIS, SMC and rejection sampling.  In our main experiments we use AIS-based estimators.  

\subsubsection{IS-based estimator}

\input{algorithms/is_estimator}

The IS-based estimator is summarized in \Cref{alg:importance-sampler}, which complements the illustration in \Cref{subsec:proposal}. 


\citet{huang2023reverse} consider a similar importance weighted score estimator (see their Eq 5 in Section 3.3) where the proposal $q(\cdot \mid \xt)=\mathcal{N}(\cdot \mid \xt/ \alpha_t, \sigma_t^2 / \alpha_t^2 \mathbb{I})$ is obtained by reversing the forward noising likelihood. While the IS-based score estimator can be inefficient when $t$ is large, \citet{huang2023reverse} leverages Unadjusted Langevin Algorithm (ULA) with IS estimator for initialization; And when $t$ is close to 0, they  are able to quickly obtain rough score estimates via the IS approach. 
 
\subsubsection{AIS and SMC-based  estimator}\label{app:ais}

\input{algorithms/ais_estimator}

\Cref{alg:annealed-importance-sampler} presents an AIS-based  approach \citep{neal2001annealed} to construct the score  and marginal estimates. AIS introduces a sequence of intermediate unnormalized distributions, smoothly bridging a tractable initial proposal distribution $q(\xzero \mid \xt)$ to the denoising posterior  $\pi(\xzero \mid \xt) \propto \tilde \pi(\xzero) \mathcal{N}(x_t \mid \alpha_t \xzero, \sigma_t^2 \mathbb{I})$. 
Under a geometric annealing schedule, intermediate targets are $\nu_k(\xzero) \propto q(\xzero \mid x_t)^{1-\beta_k}
\pi(\xzero \mid \xt)^{\beta_k}$ where $\beta_k = k/\nsteps$ for $k=0, \cdots, \nsteps$.

 \Cref{alg:annealed-importance-sampler} 
 begins by drawing a set of particles from the proposal distribution $q(\xzero \mid x_t)$. Then it gradually transitions these particles by running MCMC with a kernel invariant to the intermediate targets $\nu_{k+1}(\xzero)$ for $k=0,\cdots, \nsteps-1$. In this work we mainly consider  Metropolis-Hasting adjusted Langevin dynamics \citep[MALA,][]{roberts1996exponential} and  Hamiltonian Monte Carlo \citep[HMC,][]{neal2012mcmc}.  At each intermediate step, the algorithm updates the importance weights of each particle based on the ratio of successive intermediate targets.

 The final weighted set of particles provides a consistent estimate of the desired posterior, which is utilized to construct a score estimate $s(\xt, \ut)$, where $\ut$ denotes all the randomness in this procedure. 
 Additionally, these weights provide an unbiased estimate of marginal density $\hat \pi(\xt, \ut)$, i.e. $\mathbb{E}_{q(\ut \mid \xt)}\left[\hat \pi(\xt, \ut) \right]=Z \pi(\xt)$, where $q(\xt \mid \xt)$ is the sampling distribution of $\ut$ given $\xt$.

\paragraph{SMC adaption of \Cref{alg:annealed-importance-sampler}.} 
We can adapt \Cref{alg:annealed-importance-sampler} to an SMC-based procedure  by incorporating the resampling mechanism. In this case, the  weight update in \Cref{alg:ais:line:weight-update}  is modified to  the incremental weight computation: 
\[w^{(k,m)} \gets \frac{\nu_{k+1}(\utn{(m)})}{\nu_k(\utn{(m)})} \]
and the initial weight in \Cref{alg:ais:line:initial-weight} is denoted by $w^{(1,m)}$, for $m=1,\cdots,M$. With this adjustment, we resample particles $\{\utn{(m)}\}_{m=1}^M$ in the beginning of each iteration $k$ based on the previous weights $\{w^{(k-1,m)}\}_{m=1}^M$  for $k=2,\cdots, n$. The marginal estimate in \Cref{alg:ais:line:marginal} becomes 
\[\hat \pi(\xt, \ut) \gets \frac{1}{\nsteps} \sum_{k=1}^{\nsteps} \frac{1}{\nis} \sum_{m=1}^{\nis} w^{(k,m)}.\] 
The other aspect of \Cref{alg:annealed-importance-sampler} remains the same. 

\subsubsection{Rejection sampling (RS)-based estimator}
\input{algorithms/rejection_sampling_estimator}

In \Cref{alg:rejection-sampler}, we present a rejection sampling (RS)-based approach, building on prior work of \citet{he2024zeroth} which   propose an RS-based score estimator and develop theoretical guarantees for the resulting diffusion MC sampler. 
We extend their approach to additionally provide an unbiased marginal density estimate alongside the original score estimate. In principle, this procedure can be used within \ourmethod{} (\Cref{alg:main}) as a replacement for the AIS scheme (\Cref{alg:annealed-importance-sampler}). Future work may explore this direction particularly in low-dimensional settings, as \citet{he2024zeroth} demonstrate advantages over other score estimators  in a non-SMC context.

Notably, rejection sampling requires access to a constant $C$ that  upper bounds  the ratio of the unnormalized target to the proposal. \citet{he2024zeroth} use Newton’s method to find a proxy of $C$.

\subsubsection{Proposal choices in IS/AIS/SMC/RS-based score estimators.}
\Cref{alg:importance-sampler,alg:annealed-importance-sampler,alg:rejection-sampler} require specifying an (initial) proposal distribution $q(\xzero \mid \xt)$. The choice of this proposal is flexible and should ideally approximate the true posterior $\pi(\xzero \mid \xt)$; the closer it is, the more effective the resulting estimator.

In \Cref{alg:importance-sampler,alg:annealed-importance-sampler,alg:rejection-sampler}, the default initial proposal is obtained by reversing the forward noising likelihood, suggested by  prior work \citep{huang2023reverse, akhound2024iterated, he2024training}. This proposal is effective when the likelihood signal is strong—i.e., when $t$ is close to $0$—but becomes less efficient as $t$ increases and the signal weakens.

Alternative strategies include using a Gaussian proposal centered at the current $\xt$. If additional information about the target distribution is available—such as its mean or covariance—it can be incorporated into the proposal. For instance, one may use a Laplace approximation of the target to construct an approximate posterior, which can be used as a sensible proposal.

\subsubsection{Score estimator based on other score identities}

In  \Cref{eq:score-is-estimate} of the main paper, we derive the score estimator based on the DSI in \Cref{eq:dsi}. Alternative score estimators can be derived based on other score identities. 

Based on TSI in \Cref{eq:tsi}, we obtain another estimate: 
\begin{align}\label{eq:score-tsi-estimate}
 \nabla_{\xt}  \log p(\xt) & \approx   s(\xt, \ut) := \sum_{m=1}^M \frac{w^{(m)}}{\sum_{m'=1}^M w^{(m')}} \cdot \frac{\nabla_{\ut}\log q(\utsupm)}{\alpha_t},
\end{align}
where $\{w^{(m)}, \ut^{(m)}\}_{m=1}^M$ is a weighted system approximating the posterior $\pi(\xzero \mid \xt)$ using an IS/AIS/SMC  procedure.

Or, given MSI in \Cref{eq:msi}, we have:
\begin{align}\label{eq:score-msi-estimate}
 \nabla_{\xt}  \log p(\xt) & \approx   s(\xt, \ut) := \sum_{m=1}^M \frac{w^{(m)}}{\sum_{m'=1}^M w^{(m')}} \cdot \Big(\alpha_t \big(\utsupm + \nabla_{\ut}\log q(\utsupm) \big) -\xt \Big).
\end{align}

We primarily consider DSI and TSI-based estimators in our experiments.

\subsection{Computational complexity}\label{app:subsec:method:complexity}

The total computational cost of \ourmethod{} scales linearly with the number of diffusion discretization steps $T$, the number of particles $N$, and the cost of each score estimation step. 

Specifically, for AIS-based score estimator as described in \Cref{alg:annealed-importance-sampler}, the score estimation cost scales linearly with the number of importance samples $\nis$, the number of AIS transition steps $\nsteps$, and the number of MCMC steps per transition $\msteps$. The vanilla IS-based score estimator in \Cref{alg:importance-sampler} is a special case with $\nsteps=1$ and $\msteps=1$. Therefore  the cost of score estimation per SMC particle per time step $ \bigO(M  \nsteps \msteps)$. Consequently, the overall  complexity for RDSMC is $\bigO(NT \cdot M \nsteps \msteps )$. 

Among these factors, the operations involving discretization steps $T$, AIS transition steps $\nsteps$, and MCMC steps $\msteps$ must be processed sequentially due to their time-dependent structure. In contrast, computations over the particles $N$ and the importance samples $\nis$ are embarrassingly parallel and can be efficiently distributed across parallel compute resources.

\subsection{Implementation guidelines}\label{app:subsec:method:implement}
We discuss  practical implementation guidelines for \ourmethod{}, including the resampling strategy in \Cref{app:subsubsec:method:resampling}, score clipping in \Cref{app:subsubsec:method:score-clipping} and  hyperparameter choices in \Cref{app:subsubsec:method:hyperparam}.

\subsubsection{SMC resampling strategy}\label{app:subsubsec:method:resampling}
While the resampling step in \ourmethod{} is crucial for steering particles toward high target probability region, it introduces extra variance in the short term. We discuss several strategies to reduce this added variance. 

First, we use systematic resampling, which introduces correlations among resampling indices 
\citep[Chapter 2.2.1]{naesseth2019elements}. 

We also consider start resampling at a later stage of the sampling trajectory. 
At early steps, the Gaussian likelihood provides weaker conditional information, making posterior inference more challenging \citep{huang2023reverse,grenioux2024stochastic}. Consequently, both the score estimates and marginal density estimates can be significantly biased or exhibit high variance, especially under limited compute budgets. 
As a result, the intermediate target distributions in early steps may not be sufficiently informative to enable effective resampling. To mitigate this issue, we delay resampling until a designated step $\tstart<T$, which we treat as a tunable hyperparameter.

For timestep $t < \tstart$, we adopt an effective sample size (ESS)-based criterion to decide whether resampling should be performed at each step -- a common strategy in SMC literature \citep[Chapter 2.2.2]{naesseth2019elements}. The ESS at step $t$ is computed as 
\begin{align}\label{eq:ess}
ESS_t &:= \frac{\left(\sum_{i=1}^N \wt^{(i)}\right)^2}{\sum_{i=1}^N \left(\wt^{(i)}\right)^2}.
\end{align}

We trigger resampling only when  the normalized ESS, computed as $ESS_t/N$,is below a threshold $\essthreshold$, another hyperparameter which we fix at 0.3 throughout experiments.

When resampling is skipped (i.e. for $t> \tstart$ or when the normalized ESS is above $\essthreshold$), we must adjust the computation of the intermediate weights to ensure the unbiasedness of the log normalization constant estimate \citep[Chapter 2.2.3]{naesseth2019elements}.  For iterations where resampling is not performed, the weight update in \Cref{algline:weight} of \Cref{alg:main} is replaced by 
\begin{align}
    \wt^{(i)} &\gets \frac{\bar w_{t+1}^{(i)}}{1/N} \cdot \frac{\hat \pi_t^{(i)} p(\xtp^{(i)} \mid \xt^{(i)})}{\hat \pi_{t+1}^{(i)} q(\xt^{(i)} \mid \xtp^{(i)}, \utp^{(i)})} 
\end{align}
where $\{\bar w_{t+1}^{(i)}\}_{i=1}^N$ are the normalized weights in the previous iteration. 

\subsubsection{Score clipping}\label{app:subsubsec:method:score-clipping}
To improve numerical stability, one may consider clipping the score estimates within a pre-specified range. Note that score clipping does not affect our theoretical guarantees provided that regularity conditions hold, as the score estimate is only part of the proposal mechanism. 

As observed in prior work \citep{noble2024learned}, score estimates produced by the TSI estimator defined in \Cref{eq:tsi} can exhibit high variance or have large magnitudes, especially during early steps.  \citet{akhound2024iterated} apply score norm clipping when training diffusion samplers where the loss  function is based on the TSI. Inspired by their work, we also clip the norm of the score estimates to a maximum of 20 when we use the TSI-based score estimates. In contrast, we do not apply clipping when using the DSI-based score estimator.

\subsubsection{Hyperparameter choices}\label{app:subsubsec:method:hyperparam}

The main hyperparameters of \ourmethod{} are: 

\paragraph{Diffusion-related hyperparameters.}
\begin{itemize}
    \item Diffusion parameterization: we use the VP diffusion schedule described in \Cref{eq:vp}  in this work.
    \item Time discretization strategy:  we use a uniform time discretization grid with $T=100$ steps, though more informative strategies can be incorporated—such as the SNR-adapted scheme proposed by \citet{grenioux2024stochastic}.
\end{itemize}

\paragraph{MC score estimation-related hyperparameters.} 
\begin{itemize}
   \item Score estimator type: we use AIS-based score estimator (\Cref{alg:annealed-importance-sampler}), with vanilla IS (\Cref{alg:importance-sampler}) as a special case. We mainly leverage the DSI (\Cref{eq:score-is-estimate}) to estimate the score in most experiments, while we also use the TSI (\Cref{eq:score-tsi-estimate}) in Bayesian logistic regression tasks. 
    \item MCMC transition kernel: 
     MALA or HMC.
    \item $\nis$:  Number of importance samples used in Monte Carlo score estimation
    \item $\nsteps$: Number of intermediate distributions, i.e. number of AIS steps; for IS $\nsteps=1$
    \item $\stepsize$: 
   Step size used in the MCMC transition kernel (not used in IS). We manually tune this parameter in the main experiments while we use adaptive step size in ablation studies; see discussion below. 
    \item $\msteps$
   : Number of MCMC steps per AIS transition (not used in IS).  We use $\msteps=1$ in all experiments. 
\end{itemize}

\paragraph{SMC-related hyperparameters.} 
\begin{itemize}
    \item \( \tstart \): Starting time for resampling within the SMC trajectory. No resampling is performed for \( t/T > \tstart \).
    \item \( \essthreshold \): Normalized effective sample size threshold triggering resampling. When \( \essthreshold = 0 \), no resampling is applied throughout, which corresponds to  \ourmethod{} (IS) if a final-step weighting is applied and  \ourmethod{} (Proposal) otherwise. We set \( \essthreshold = 0.3 \) in all experiments for \ourmethod{}. 
\end{itemize}

In the main experiments, we choose score estimator and MCMC kernel types by a coarse grid search, and tune $\nis, \nsteps, \stepsize$ and $\tstart$ in a range of values,  using validation metrics. The hyperparameter values are summarized \Cref{tab:hyperparam-grid-our};  see  details in \Cref{app:subsubsec:hyperparameters}. 

In the ablation experiments in \Cref{app:subsec:ablation}, we fix most hyperparameters to reduce the degrees of freedom. We use $T=100, M=100, n=50, m=1, \essthreshold=0.3$, and AIS-based score estimator leveraging the DSI,  LG as the MCMC kernel. We also use an \textit{adaptive} MCMC step size $\stepsize$, as in  \citet{grenioux2024stochastic}. When the  acceptance rate is too high ($>  76\%$) we increase the step size by $3\%$, when it is $< 74\%$, we decrease step size by $3\%$, and otherwise we keep the current step size.  Only  an initial step size value needs to be specified and then it is adpatively adjusted during the sampling process.  The \textit{only}  hyperparameter in \ourmethod{} that requires tuning is $\tstart$ , which we vary in $\{0, 0.1,\cdots, 0.9,1.0\}$. We observe consistent performance and  therefore in practice we would recommend users to start with something similar to this configuration.

%% file: algorithms/is_estimator.tex
\begin{algorithm}[!t]
\caption{Importance Sampling (IS)-based score and marginal estimator}
\label{alg:importance-sampler}
\KwIn{Unnormalized prior $ \tilde \pi(\cdot)$, likelihood parameters $\{\alpha_t ,\sigma_t\}$, data $\xt$, number of importance samples  $M$, and proposal distribution $q(\cdot \mid x_t)$ (default: $\mathcal{N}(\cdot \mid \xt/ \alpha_t, \sigma_t^2 / \alpha_t^2 \mathbb{I})$)} 
\KwOut{Score estimate $s(\xt, \ut)$ and marginal estimate $\hat \pi(\xt, \ut)$.}
\For{$m \gets 1$ \KwTo $M$}{
Sample $\utn{(m)} \sim q(\cdot \mid \xt)$\\
Compute importance weights  $w^{(m)} \gets \frac{\tilde \pi(\utsupm) \mathcal{N}(\xt \mid \alpha_t \utsupm, \sigma_t^2I)}{q(\utsupm)}$\\
}
Denote auxiliary variable $\ut \gets \{\utn{(m)}\}_{m=1}^M$\\ 
Compute score estimate  $s(\xt, \ut) \gets \sum_{m=1}^M \frac{w^{(m)}}{\sum_{m'=1}^M w^{(m')}} \cdot \frac{\alpha_t \utsupm -\xt}{\sigma_t^2} 
$ \\
Compute marginal estimate  $\hat \pi(\xt, \ut) \gets \frac{1}{M} \sum_{m=1}^M w^{(m)} $.
\end{algorithm}

%% file: algorithms/ais_estimator.tex
\begin{algorithm}[!t]
\caption{Annealed Importance Sampling (AIS)-based Score and Marginal Estimator}
\label{alg:annealed-importance-sampler}
\KwIn{Unnormalized prior $\tilde \pi(\cdot)$, likelihood parameters $\{\alpha_t, \sigma_t\}$, data $x_t$, number of  importance samples $M$,   number of annealing steps $n$, initial proposal distribution $q(\cdot \mid x_t)$ (default: $\mathcal{N}(\cdot \mid \xt/ \alpha_t, \sigma_t^2 / \alpha_t^2 \mathbb{I})$), and MCMC-related inputs \{number of MCMC steps per transition $\msteps$,  MCMC transition kernel $\mathcal{T}$ and MCMC step size $\stepsize$\}}
\KwOut{Score estimate $s(\xt, \ut)$ and marginal estimate $\hat \pi(\xt, \ut)$}

Set annealing schedule $\beta_k \gets k/\nsteps$ for $k=0, \dots, \nsteps$ \\
Set initial target 
        $\nu_1(\xzero) \gets q(\xzero \mid \xt)^{1-\beta_1}\tilde \pi(\xzero)^{\beta_1}\mathcal{N}(\xt \mid \alpha_t \xzero, \sigma_t^2 \mathbb{I})^{\beta_1}$ \\
\For{$m \gets 1$ \KwTo $M$}{
    Draw initial sample $\utn{(1,m)} \sim q(\cdot \mid x_t)$ \\
     Initialize weight $w^{(m)} \gets \frac{\nu_1(\utn{(1,m)})}{\nu_0(\utn{(1,m)})}$ \label{alg:ais:line:initial-weight}\\
}    
\For{$k \gets 2$ \KwTo $\nsteps $}{
Set the current target 
$\nu_{k}(\xzero) \propto q(\xzero \mid \xt)^{1-\beta_{k}}\tilde \pi(\xzero)^{\beta_{k}}\mathcal{N}(\xt \mid \alpha_t \xzero, \sigma_t^2 \mathbb{I})^{\beta_{k}}$ \\
\For{$m \gets 1$ \KwTo $\nis$}{
Update sample $\utn{(k,m)}$ using MCMC transition  targeting $\nu_{k}$ starting from $\utn{(k-1,m)}$\\ 
        Update weight 
        $ w^{(m)} \gets w^{(m)} \frac{\nu_{k}(\utn{(k,m)})}{\nu_{k-1}(\utn{(k,m)})} $\\ \label{alg:ais:line:weight-update}
    }
}
Denote auxiliary variable $\ut \gets \{\{\utn{(k,m)} \}_{m=1}^\nis\}_{k=1}^{\nsteps}$ \\
Compute normalized weights  $\tilde w^{(m)} \gets \frac{w^{(m)}}{\sum_{m'=1}^{\nis} w^{(m')}}$ \\
Compute score estimate $s(\xt, \ut) \gets \sum_{m=1}^{\nis} \tilde w^{(m)} \frac{\alpha_t \utn{(k,m)} - x_t}{\sigma_t^2}$ \\
Compute marginal density estimate $\hat \pi(\xt, \ut) \gets \frac{1}{\nis} \sum_{m=1}^{\nis} w^{(m)}$ \label{alg:ais:line:marginal}
\end{algorithm}

%% file: algorithms/rejection_sampling_estimator.tex
\begin{algorithm}[!t]
\caption{Rejection Sampling (RS)--based score and marginal estimator}
\label{alg:rejection-sampler}
\KwIn{Unnormalized prior $\tilde{\pi}(\cdot)$, likelihood parameters $\{\alpha_t, \sigma_t\}$, data $\xt$, number of accepted samples $M$, proposal $q(\cdot \mid \xt)$ (default: $\mathcal{N}\left(\cdot \mid \xt /\alpha_t, \alpha_t^2 / \sigma_t^2 \mathbb{I}\right)$), and an upper bound  $C \ge \sup_{\ut} \tilde{\pi}(\ut) \, \mathcal{N}(\xt \mid \alpha_t \ut, \sigma_t^2 I) / q(\ut)$}
\KwOut{Score estimate $s (\xt, \ut)$ and marginal estimate $\hat{\pi}(\xt, \ut)$.}

$m, i \gets 1, 1$ \\
\While{$m \leq M$}{
    Sample proposal $\nui \sim q(\cdot \mid \xt)$\\
    Compute acceptance ratio $r \gets \frac{\tilde{\pi}(\nui) \, \mathcal{N}(\xt \mid \alpha_t \nui, \sigma_t^2 I)}{C \, q(\nui)}$\\
    Draw $\vi \sim \mathrm{Uniform}(0,1)$\\
    \If{$v < r$}{
        Accept sample $\utn{(m)} \gets \nui$\\ 
        $m \gets m+1$\\
    }
    $i \gets i+1$\\
}
$M_\textrm{total} \gets i-1$\\
Denote auxiliary variable $\ut \gets \{ \nui, \vi\}_{i=1}^{M_\textrm{total}}$\\
Compute score estimate: $s(\xt, \ut) \gets \frac{1}{M} \sum_{m=1}^M \frac{\alpha_t \utn{(m)} - \xt}{\sigma_t^2}$\\
Compute marginal estimate:  $\hat{\pi}(\xt, \ut) \gets C \frac{M}{M_{\textrm{total}}}$\\
\end{algorithm}

%% file: sections/appendix_theory.tex
\section{Theoretical guarantees of \ourmethod{}}\label{app:sec:theory}
We formally state the  theorem providing sufficient conditions for asymptotic accuracy of \ourmethod{} and unbiasedness of the estimate of the normalization constant. Our proof strategy for asymptotic accuracy is adapted from \citet{wu2023practical} and the proof for unbiasedness follows the techniques presented in \citet{naesseth2019elements}.

\begin{theorem}\label{thm:main}
Assume
\begin{enumerate}[label=(\alph*)]
    \item{the first  marginal estimate $\hat \pi(\xT,\uT),$ and the ratios of subsequent ratio $\hat \pi(\xt, \ut) / \hat \pi(\xtp,\utp) $ are positive and bounded,}\label{item:twist_assumption}
    \item{the score estimate $s (\xt,\ut)$ is bounded, and $\xt$ has compact support,} \label{item:twist_gradient}
    \item{variance increases in the forward noising kernel, i.e.\ for each $t,\ \gtp^2 >\gt^2.$} \label{item:proposal_var}
\end{enumerate}
Then the \ourmethod{} algorithm described in \Cref{alg:main} provides a consistent estimator of the target distribution $\pi(x_0)$ as particle size  $N\to\infty$ and an unbiased estimator of the normalization constant $Z$ for any $N\geq 1$. More specifically, 
\begin{enumerate}[label=(\alph*)]
    \item Let $\mathbb{P}_{N}=\sum_{i=1}^N w_0^{(i)}\delta_{\xn{0}^{(i)}}$ for weighted particles $\{(\xn{0}^{(i)}, w_0^{(i)})\}_{i=1}^N$ returned by \Cref{alg:main} with $N$ particles. $\mathbb{P}_{N}$ converges setwise to $\pi(\xzero)$ with probability one, that is for every set $A,$ $\lim_{N\rightarrow\infty} \mathbb{P}_{N}(A) = \int_A \pi(\xzero) d\xzero.$
    \item For the estimate for the normalization constant $\hat Z$ returned by \Cref{alg:main}, we have $\mathbb E \hat Z = Z$.
\end{enumerate}
\end{theorem}

The assumptions of \Cref{thm:main} are typically satisfied in standard implementations of SMC samplers:
\begin{itemize}
    \item{Assumption \ref{item:twist_assumption} can be satisfied by the clipping of the  marginal estimates $\hat \pi(\xt, \ut)$ within the range $[c,C]$ for some constants $c, C > 0$.}
    \item{Assumption \ref{item:twist_gradient} by clipping scores estimates as well. Additionally, the compactness of the state space \( \xt \) is a standard and commonly adopted assumption in the SMC literature.}

    \item{Assumption \ref{item:proposal_var} is mild and typically satisfied in practice. In both VE and VP noising schedules commonly used in diffusion models, the variance of the forward kernel increases with time.}
\end{itemize}

\paragraph{Proof of \Cref{thm:main}:}

We first prove the asymptotic exactness of our sampler and then prove the unbiasedness of the normalization constant estimate.

First, \Cref{thm:chopin_theorem} characterizes a set of conditions under which SMC algorithms converge.
We restate this result below in our own notation.
\begin{theorem}[\citet{chopin2020introduction} -- Proposition 11.4]\label{thm:chopin_theorem}
Let $\{(\xn{0}^{(i)}, w_0^{(i)})\}_{i=1}^N$ be the particles and weights returned at the last iteration of a sequential Monte Carlo algorithm with $N$ particles using multinomial resampling.
If each weighting function $w_t$ is positive and bounded, 
then for every bounded, $\nu_0$- measurable function $\phi$ of $\xt$
$$
\lim_{N\rightarrow \infty} \sum_{i=1}^N w_0^{(i)} \phi(\xn{0}^{(i)})
=
\int \phi(\xzero)\nu_0(\xzero)d\xzero.
$$
with probability one.
\end{theorem}

An immediate consequence of \Cref{thm:chopin_theorem} is the setwise convergence of the discrete measures. To apply \Cref{thm:chopin_theorem}to show asymptotic accuracy, it is sufficient to show that the weights at each step are upper bounded, as they are defined through probabilities and hence are positive. Since there is a finite number of steps $T$, it suffices to show that each $w_t$ is bounded.

The initial weight is defined by
\[
w_T = \frac{\hat \pi(x_T,\ut)}{q(x_T)},
\]
which is bounded by Assumption~\ref{item:twist_assumption}.

The intermediate weights are defined by \Cref{eq:weight} such that 
\[
w_t = \frac{\hat \pi(x_t,\ut)\,\pi(\xtp\mid x_t)}{\hat \pi(\xtp,\utp)\,q(x_t\mid \xtp,\utp)}.
\]
We first decompose the weights by 
\[
\log w_t = \log \frac{\hat \pi(x_t,\ut)}{\hat \pi(\xtp,\utp)} + \log \frac{\pi(\xtp\mid x_t)}{q(x_t\mid \xtp,\utp)}.
\]

The boundedness of $\log \frac{\hat \pi(x_t,\ut)}{\hat \pi(\xtp,\utp)}$ is guaranteed by Assumption~\ref{item:twist_assumption}. We now show the boundedness of $\log \frac{\pi(\xtp\mid x_t)}{q(x_t\mid \xtp,\utp)}$. First, note that 
\[
\pi(\xtp\mid x_t) = \mathcal{N}(\xtp \mid \xt + \ft \xt \delta , \gt^2 \delta),
\]
and
\[
q(\xt \mid \xtp, \utp) = \mathcal{N}(\xt \mid \xtp - \left[ \ftp \xtp - \gtp^2   s(\xtp, \utp)\right] \delta , \gtp^2 \delta).
\]

Let 
$
\hat \mu_p = \xtp - \ft \xt \delta \ \text{ and } \ \hat \mu_q = \xtp - \left[ \ftp \xtp - \gtp^2   s(\xtp, \utp)\right] \delta. 
$
We can show that $\|\hat\mu_p - \hat \mu_q\|  = \| \ftp \xtp - \ft \xt - \gtp^2   s(\xtp, \utp)\| \delta$ is bounded , which follows from Assumption \ref{item:twist_gradient}. The log-ratio then simplifies as 
\begin{align}
&\log \frac{\pi(\xtp\mid x_t)}{q(\xt \mid \xtp, \utp)} \\
&= 
\log \frac{| 2\pi \gt^2 I|^{-1/2} \exp\{-(2 \gt^2)^{-1}\|\xt - \hat \mu_p\|^2 \}}
{ | 2\pi \gtp^2 I|^{-1/2} \exp\{-(2 \gtp^2)^{-1}\|\xt -\hat \mu_q\|^2 \}} \\
&\textrm{Rearrange \ and\  let  \ } C=\log | 2\pi \gt^2 I|^{-1/2} /  | 2\pi \gtp^2  I|^{-1/2}\\
&=  - \frac{1}{2}\left[ 
\gt^{-2}\|\xt - \hat \mu_p \|^2    -
\gtp^{-2}\|\xt -\hat \mu_q \|^2 \right]  +C \\
&\textrm{Expand and rearrange \ } \|\xt -\hat \mu_q \|^2 = \|\xt - \hat \mu_p \|^2 + 2 \langle \hat\mu_p - \hat \mu_q, \xt - \hat \mu_p  \rangle + \|\hat\mu_p - \hat \mu_q\|^2 \\
&= - \frac{1}{2}\Big[(\gt^{-2} - \gtp^{-2})\|\xt - \hat \mu_p \|^2  - 2
\gtp^{-2}\langle \hat\mu_p - \hat \mu_q, \xt - \hat \mu_p  \rangle - \gtp^{-2}\|\hat\mu_p - \hat \mu_q\|^2  \Big] + C \\
&\textrm{Let \ } C^\prime {=} C + \frac{1}{2} \gtp^2 \|\hat\mu_p - \hat \mu_q\|^2 \textrm{\ and rearrange.  Note that $\|\hat\mu_p - \hat \mu_q\|^2{<}\infty$}\\
&= - \frac{1}{2}(\gt^{-2} - \gtp^{-2})\|\xt - \hat \mu_p \|^2 +
\gtp^{-2}\langle \hat\mu_p - \hat \mu_q, \xt - \hat \mu_p  \rangle + C^\prime \\
&\textrm{By Cauchy-Schwarz, \ } \\
& \leq - \frac{1}{2}(\gt^{-2} - \gtp^{-2})\|\xt - \hat \mu_p \|^2 +
\gtp^{-2}\|\hat\mu_p - \hat \mu_q \| \cdot \| \xt - \hat \mu_p  \| + C^\prime \\
&\textrm{By Assumption \ref{item:proposal_var}}, \gt^{-2} - \gtp^{-2}>0, \\
& \leq \frac{1}{2} \frac{(\gtp^{-2}\|\hat\mu_p - \hat \mu_q \|)^2}{\gt^{-2} - \gtp^{-2}} + C^\prime \\
& = \frac{\gtp^{-4}}{2(\gt^{-2} - \gtp^{-2})}\|\hat\mu_p - \hat \mu_q \|^2 + C^\prime \\
&\le C^{\prime\prime}.
\end{align}
The final line follows from the boundedness of $\|\hat\mu_p - \hat \mu_q \|$.
The above derivation shows that each $w_t$ is bounded, which concludes the proof for the asymptotic exactness of our sampler.

Then we show the unbiasedness of the normalization constant estimate $\hat Z$. Let $a^i_t$ be the index of $\xt^{(i)}$ and $\utn{(i)}$ before resampling. With that, $\hat Z$ is defined by
\[
\hat Z = \prod_{t=0}^T \frac{1}{N}\sum_{i=1}^N \wt^{(i)},
\quad \text{where } 
\wt^{(i)} = \frac{\hat \pi_t^{(i)} \pi(\xtp^{(i)} \mid \xt^{a^i_t})}{\hat \pi_{t+1}^{(i)} q(\xt^{a^i_t} \mid \xtp^{(i)}, \utpn{(i)})}, \ \wT^{(i)} = \frac{\hat \pi_T^{(i)}}{\hat \pi(\xT^{a^i_t}) }.
\]

Denote the normalized weight by $\widetilde \wt^{(i)} = \frac{\wt^{(i)}}{\sum_{i=1}^N \wt^{(i)}}$. The distribution of all random variables generated by the SMC method is then given by
\begin{align*}
    & Q(\xn{0:T}, \un{0:T}, a_{0:T}) \\
    &= \pi(\xt, \uT) p(\xn{0:{T-1}}, \un{0:{T-1}}, a_{0:T}) \\
    &= \pi(\xt^{a_T}, \uT^{a_T}) \pi(\xt, \uT \mid \xT^{a_T}, \uT^{a_T}) \prod_{t=0}^{T-1} p(\xn{t}, \un{t} \mid \xn{t}^{a_t}, \un{t}^{a_t}) q(\xn{t}^{a_t} \mid \xtp, \utp) q(\un{t}^{a_t} \mid \xn{t}^{a_t}), \\
    &= \pi(\xt^{a_T}) \Big( \prod_{i=1}^N \frac{\wT^{(i)}}{\sum_{i=1}^N \wT^{(i)}} \Big)\prod_{i=1}^N \Big[  \Big(\prod_{t=0}^{T-1}  \frac{\wt^{(i)}}{\sum_{i=1}^N \wt^{(i)}} q(\xt^{a_t^i}  \mid \xtp^{(i)}, \utpn{(i)}) \Big)\Big]\\
    & \Big[ q(\uT^{a_T}\mid \xT^{a_T}) \prod_{t=0}^{T-1} q(\un{t}^{a_t} \mid \xn{t}^{a_t}) \Big] .
\end{align*}

By $w_t^{(i)} = \frac{\hat \pi_t^{(i)} \pi(\xtp^{(i)}|\xt^{a_t^i})}{\hat \pi_{t+1}^{(i)} q(\xt^{a_t^i}  \mid \xtp^{(i)}, \utpn{(i)}) }$ and $w_T^{(i)} = \frac{\hat \pi_T^{(i)}}{q(x_T^{a_T^i})}$, we have:
\begin{align*}
     & Q(\xn{0:T}, \un{0:T}, a_{0:T}) \\
     &= \Big(\prod_{t=0}^{T}\sum_{i=1}^N \wt^{(i)}\Big)^{-1} \hat \pi_0^{(1)} \Big
     (\prod_{t=0}^{T-1}  \pi(\xtp^{(1)}|\xt^{a_t^1})\Big) \\
     & \prod_{i=2}^N \hat \pi_T^{(i)} \prod_{i=2}^N \Big[  \Big(\prod_{t=0}^{T-1}  \widetilde \wt^{(i)} q(\xt^{a_t^i}  \mid \xtp^{(i)}, \utpn{(i)}) \Big)\Big]
    \Big[ q(\uT^{a_T}\mid \xT^{a_T}) \prod_{t=0}^{T-1} q(\un{t}^{a_t} \mid \xn{t}^{a_t}) \Big] ,
\end{align*}
where we plug in $w_t^{(1)}$ and rearrange terms.

By multiplying $\hat Z$, we have:
\begin{align*}
    \hat Z Q(\xn{0:T}, \un{0:T}, a_{0:T}) &=  \hat Z \Big(\prod_{t=0}^{T}\sum_{i=1}^N \wt^{(i)}\Big)^{-1} \hat \pi_0^{(1)} \Big (\prod_{t=0}^{T-1}  \pi(\xtp^{(1)}|\xt^{a_t^1})\Big)   \\
   & \prod_{i=2}^N \Big[  \hat \pi_T^{(i)} \Big(\prod_{t=0}^{T-1}  \widetilde \wt^{(i)} q(\xt^{a_t^i}  \mid \xtp^{(i)}, \utpn{(i)}) \Big)\Big] \Big[ q(\uT^{a_T}\mid \xT^{a_T}) \prod_{t=0}^{T-1} q(\un{t}^{a_t} \mid \xn{t}^{a_t}) \Big] \\
    &= N^{-T} \hat \pi_0^{(1)} q(u_0^{a_0^1}\mid x_0^{a_0^1}) \Big (\prod_{t=0}^{T-1}  \pi(\xtp^{(1)}|\xt^{a_t^1}) q(\utp^{a_{t+1}^1}\mid \xtp^{a_{t+1}^1})\Big)  \\
    & \prod_{i=2}^N \Big[  \hat \pi_T^{(i)} q(u_t^{a^i_T}\mid x_t^{a^i_T}) \Big(\prod_{t=0}^{T-1}  \widetilde \wt^{(i)} q(\xt^{a_t^i}  \mid \xtp^{(i)}, \utpn{(i)}) q(\utp^{a_{t+1}^{(i)}}\mid \xtp^{a_{t+1}^{(i)}}) \Big)\Big],
\end{align*}
where the first equality follows from plugging in $Q$ and the second follows from plugging in $\hat Z$.

The expectation of the normalization constant estimate can be written as
\begin{align*}
    \mathbb E_Q(\hat Z) =& N^{-T} \sum_{{a_{0:T}}} \int  \int \hat \pi_0^{(1)} q(u_0^{a_0^1}\mid x_0^{a_0^1}) \Big (\prod_{t=0}^{T-1}  \pi(\xtp^{(1)}|\xt^{a_t^1}) q(\utp^{a_{t+1}^1}\mid \xtp^{a_{t+1}^1})\Big) \\
    & \prod_{i=2}^N \Big[  \hat \pi_T^{(i)} q(u_t^{a^i_T}\mid x_t^{a^i_T}) \Big(\prod_{t=0}^{T-1}  \widetilde \wt^{(i)} q(\xt^{a_t^i}  \mid \xtp^{(i)}, \utpn{(i)}) q(\utp^{a_{t+1}^{(i)}}\mid \xtp^{a_{t+1}^{(i)}}) \Big)\Big] d\xn{0:T} d\un{0:T} \\
    = & N^{-T} \sum_{{a_{0:T}^1}} \int  \int \hat \pi_0^{(1)} q(u_0^{a_0^1}\mid x_0^{a_0^1}) \Big (\prod_{t=0}^{T-1}  \pi(\xtp^{(1)}|\xt^{a_t^1}) q(\utp^{a_{t+1}^1}\mid \xtp^{a_{t+1}^1})\Big) d x_{0:T}^1 d u_{0:T}^1 \\
    = & Z,
\end{align*}
where in the second equality, we marginalize the variables not involved in the particle $x_{0:T}^1$ and $x_{0:T}^{a_{0:T}^1}$. The final equality follows because we are averaging over $N^T$ possible cases of $a_{0:T}^1$ and all are equal to Z, which concludes the proof.

%% file: sections/appendix_related_works.tex
\section{Comparison to Twisted Diffusion Sampler}\label{app:sec:related-works}

In this section, we discuss similarities and differences between \ourmethod{} and  Twisted Diffusion Sampler \citep[TDS,][]{wu2023practical}, which is an SMC method for conditional generation from diffusion models. 

Structurally, our method shares similarities with TDS, as both methods use some form of "look-ahead" or "twisting" functions \citep[][Chapter 3]{naesseth2019elements} to improve the efficiency of SMC. These twisting functions introduce future information to intermediate states, promote promising particles through the weighting procedure, and improve the SMC sampling efficiency.

Moreover, the incremental weight expression of both TDS and RDSMC takes the same form (their Eq. (11) v.s. our \Cref{eq:weight-final}):
\begin{align*}
    w_t &:= \frac{\psi(\xt) \gamma^0_t(\xn{t:T})}{\psi(\xtp) \gamma^0_t(\xn{t+1:T}) q(\xt \mid \xtp)},
\end{align*}
 where $\psi$ 
 is the twisting function, 
 $\gamma^0_t$ is the untwisted intermediate target and $q$ 
 is the proposal. The product $\gamma_t(\xn{t:T}) \coloneqq \psi(\xt) \gamma_t^0 (\xn{t:T})$ defines a series of \textit{twisted} intermediate targets.
  
 While the   expressions of these components are method-specific, the structure of the weighting functions is shared between TDS our \ourmethod{} and follows from the generic twisted SMC framework \citep[][Chapter 3]{naesseth2019elements}. 

Importantly, this weight construction introduces twisting functions in a telescoping manner such that their approximation errors cancel out over the full trajectory. This property ensures that the final target is correct, and the resulting SMC procedure is asymptotically exact for both TDS and RDSMC.

However, we want to highlight several important \textit{differences} between TDS and RDSMC:
\begin{enumerate}
    \item 
 Problem setting: The problem we address is fundamentally different from that of TDS. TDS is designed to sample from the conditional distribution $p_\theta (\xzero \mid y) \propto p_\theta(\xzero) p(y \mid \xzero)$ 
 given a pretrained diffusion model $p_\theta (\xzero)$
 and a likelihood model $p(y \mid \xzero)$
 with the observation $y$. As a comparison, our goal is to sample from an arbitrary distribution $\pi(\xzero)$, assuming access to its unnormalized density $\tilde \pi(\xzero)$.

 \item Design of key components: Due to the difference in settings, our intermediate target distributions, proposal distributions, and twisting functions require different designs. In particular,
\begin{itemize}
\item 
 TDS’s twisting function is 
$ p(y \mid \hat x_\theta(\xt)$
 (their Eq. (8)), the likelihood evaluated at the denoising prediction $\hat x_\theta(\xt)$ by the pretrained model, which approximates the intractable likelihood $p_\theta(y\mid \xt)$.  
 This twisting function incorporates the observation $y$ 
 into the weighting function (their Eq. (11)). Notably, when $t=0$, TDS's twisting function recovers the exact likelihood $p(y \mid \xzero)$ and no approximation is involved. 

 Given the twisting function, TDS's  intermediate target is  $p\left(y \mid \hat x_\theta (\xt) \right) p_\theta(\xn{t:T})$ where $p_\theta(\xn{t:T})$ is the distribution under the pretrained diffusion model. The proposal is obtained by adding a  term that is the gradient of the twisting function, $\nabla_{\xt} \log p(y \mid \hat x_\theta(\xt))$, to the pretrained diffusion model's transition kernel (see their Eq (9) and (11)).

\item In our case, RDSMC uses the marginal estimate $\hat \pi (\xt, \ut)$
 (\Cref{eq:marginal-estimate}) as the twisting function, which approximates the exact marginal $\pi(\xt)$. 
 The  marginal  integrates out the future information in $\pi(\xzero)$, thereby encoding the target information 
 into the intermediate weight. 

We define intermediate targets  as $\hat \pi(\xt, \ut) \pi(\xn{t:T})  q(\un{t:T} \mid \xn{t:T}) $ (a restatement of \Cref{eq:intermediate-target}); the product term $\hat \pi(\xt, \ut) \pi(\xn{t:T})$ is structurally similar to TDS's intermediate target, while we additionally account for MC estimation randomness $q(\un{t:T} \mid \xn{t:T})$. We also utilize twisting function to inform the proposal, that is,  we use the score estimate $s(\xt, \ut)$, which is connected to the twisting function / marginal estimate $\hat \pi(\xt,\ut)$,  to design the proposal in \Cref{eq:proposal}. 

 \end{itemize}

\item Use of auxiliary variables: Our method introduces auxiliary randomness through the Monte Carlo estimation of the twisting function. Incorporating this randomness places  \ourmethod{} in the category of nested SMC methods \citep{naesseth2015nested}. In contrast, TDS does not use auxiliary variables and operates more like a standard SMC algorithm, without requiring nested sampling.

\end{enumerate}

%% file: sections/appendix_experiments.tex
\section{Experiment details}\label{app:sec:experiments}
This section provides  experimental details and additional results.

\Cref{app:subsec:exp-implementation-details} describes the implementation and hyperparameter settings for \ourmethod{} and all baseline methods.
\Cref{app:subsec:gmm,app:subsec:rings-and-funnel,app:subsec:logistic-regression} present supplementary information and additional results for the target distributions used in the \textit{main} experiments. 
\Cref{app:subsec:ablation} reports  \textit{additional} experiments that control for hyperparameters, runtime, and other relevant factors to compare across methods, as well as an empirical comparison to \citet{phillips2024particle}.

\paragraph{Compute resources.} 
All experiments were conducted on a single NVIDIA A100 GPU with 40 GB  GPU memory, and  accessed between 16 GB and 32 GB of CPU memory, depending on the task. \ourmethod{} and baseline methods are all implemented in PyTorch \citep{paszke2019pytorch}.

\subsection{Method and hyperparameter details}\label{app:subsec:exp-implementation-details}

We first give an overview of baseline methods we benchmark \ourmethod{} against in \Cref{app:subsubsec:baseline-summary}. We then describe key hyperparameter settings of our method and  competing baselines for the main experiments in \Cref{app:subsubsec:hyperparameters}. 

\subsubsection{Summary of baseline methods}\label{app:subsubsec:baseline-summary}

\paragraph{AIS \citep{neal2001annealed}.} The annealed importance sampling (AIS) algorithm defines a sequence of geometrically annealed distributions $\rho_t(x)$ for $t \in \{0, 1, \cdots, T\}$ from  an initial proposal $\rho_0(x)$ to the desired target $\rho_T(x) \propto \pi(x)$ as $\rho_t (x) \propto \rho_0 (x)^{1-\beta_t} \pi(x)^{\beta_t}$ where $\{\beta_t\}_{t=0}^T$ is an increasing linear schedule with each $\beta_t = t/T$. AIS begins by sampling from the initial proposal $\rho_0(x)$ and proceeds through a sequence of MCMC transitions, each targeting the intermediate distribution $\rho_t(x)$. The resulting weighted samples provide a consistent approximation to the target distribution $\pi(x)$. 
The complexity of AIS is $\bigO(NTn)$ for $N$ final samples, $T$ annealing steps, and $n$ MCMC transitions per step.

\paragraph{SMC \citep{del2006sequential}.} Sequential Monte Carlo (SMC)  operates on the same sequence of annealed distributions as AIS but differs by performing resampling at certain iterations. In our implementation, we consider both constant resampling at each step and an adaptive strategy based on an ESS threshold $\essthreshold$. The complexity of SMC is the same as that of AIS.

\paragraph{RDMC \citep{huang2023reverse}.} 
Reverse diffusion Monte Carlo (RDMC) builds on the time‐reversed Ornstein–Uhlenbeck diffusion:
\begin{equation*}
    d Y_t = \bigl\{Y_t + 2\,\nabla\log p_{\tinit-t}(Y_t)\bigr\}\,d t + \sqrt{2}\,d B_t,\quad Y_0\sim \pi_{\tinit}
\end{equation*}
where $p_s(y)=\int \mathcal{N}(y;\,\mathrm e^{-s}x,\,(1-\mathrm e^{-2s})I)\,\pi(d x)$.  By Tweedie’s formula, the intractable score $\nabla\log p_{\tinit-t}(Y_t)$ can be written in terms of the posterior denoiser:
\begin{equation*}
u_t(y)\;=\;\int x\,q_t(x\mid y)\,d x,\qquad q_t(x\mid y)\propto\pi(x)\,\mathcal{N}\bigl(x;\,\mathrm e^{\tinit-t}y,\,(\mathrm e^{2(\tinit-t)}-1)\,I\bigr),
\end{equation*}
so that the SDE becomes
\begin{equation*}
d Y_t = \big\{\tfrac{\mathrm e^{-2\,(\tinit-t)}+1}{\mathrm e^{-2\,(\tinit-t)}-1}\,Y_t + \tfrac{2\,\mathrm e^{-(\tinit-t)}}{1-\mathrm e^{-2\,(\tinit-t)}}\,u_t(Y_t)\big\}d t + \sqrt{2}\,d B_t.
\end{equation*}

Following the implementation of \citet{grenioux2024stochastic}, RDMC discretizes the interval $[0, \tinit]$ into $T$ steps. At initialization, RDMC employs a \textit{Langevin-within-Langevin}  scheme to sample from $\pi_{\tinit}$: using  ULA where the score function is estimated based on DSI.  In subsequent steps, 
RDMC simulates from the reverse SDE where the denoiser $u_t(y)$ is obtained by running  MALA targeting $q_t(x\mid y)$.

The overall complexity of RDMC is $\bigO(NM( n_{\textrm{init}} + Tn))$ where $N$ is the number of final samples,  $T$ is the number of iterative sampling steps, and the remaining factors are associated with score estimation:  $M$ is the number of MCMC chains, while $n_{\textrm{init}}$ and $n$ denote the number of MCMC transition steps in the initial  and subsequent stages, respectively.

\paragraph{SLIPS \citep{grenioux2024stochastic}.}

Stochastic Localization via Iterative Posterior Sampling (SLIPS) relies on a stochastic observation process defined as:
\begin{equation*}
Y_t = \alpha(t) X + \sigma W_t,
\end{equation*}
where $(W_t)_{t\ge0}$ is a standard Brownian motion independent of $X \sim \pi$, and $\alpha(t)$ is a flexible denoising schedule function. To bypass the direct sampling requirement from $\pi$, SLIPS introduces a conditional denoiser defined by:
\begin{equation*}
u_t(y)=\int x q_t(x\mid y) dx, \quad q_t(x\mid y) \propto \pi(x)\mathcal{N}\left(\frac{y}{\alpha(t)}, \frac{\sigma^2}{g(t)^2}I\right).
\end{equation*}
The corresponding SDE governing this observation process is: 
\begin{align}\label{eq:slips-reverse-sde}
dY_t = \dot{\alpha}(t) u_t(Y_t) dt + \sigma dB_t,
\end{align}
where $(B_t)_{t\ge0}$ is another standard Brownian motion.

Similar to RDMC, the SLIPS algorithm approximates the dynamics defined by this SDE using an MCMC approach to estimate the conditional denoiser $u_t$ or a related score function:
\begin{itemize}
    \item The SLIPS initialization also involves a \textit{Langevin-within-Langevin} procedure. Different from RDMC, SLIPS chooses the initialization time $\tinit$ such that both the marginal $\pi_{\tinit} (y)$ and the posterior $q_{\tinit} ( x \mid y)$ are (approximately) log-concave , which they refer to as the ``duality of log-concavity" assumption.  
    
    \item In the subsequent steps, SLIPS integrates the observation process where the denoiser $u_t(Y_t)$ is estimated by MALA sampling from $q_t(x \mid y)$. 
\end{itemize}

Similar to RDMC, SLIPS' complexity is $\bigO(N M(n_\textrm{init} + Tn))$ where  $N$ is the number of final samples,  $T$ is the number of iterative sampling steps, and the remaining factors are associated with score estimation:  $M$ is the number of MCMC chains, while $n_{\textrm{init}}$ and $n$ denote the number of MCMC transition steps in the initial  and subsequent stages, respectively.

\paragraph{SMS \citep{saremi2023chain}.}
For a target \(\pi\), Sequential Multimeasurement Sampler (SMS) draws \(M\) independent noisy observations  at  some noise scale \(\sigma>0\):
\[
Y^m = X + \sigma Z^m,\quad Z^m\sim\mathcal{N}(0,I),\quad m=1,\dots,M,
\]  
with \(X\sim\pi\).  For any $m \in \{1,...,M\}$, the posterior density of \(X\) given  $y_{1:m}$ is  
\[
q_m(x\mid y_{1:m})\;\propto\;\pi(x)\,\mathcal{N}\!\bigl(\,\bar y_{1:m},\,\tfrac{\sigma^2}{m}I\bigr),
\]  
where \(\bar y_{1:m}=(1/m)\sum_{i=1}^m y_i\).  Its Bayes estimator  
\[
u_m(y_{1:m}) \;=\; \mathbb{E}[X\mid Y^{1:m}=y_{1:m}]
\]
serves as a non‐Markovian denoiser, and one shows that the law of \(u_m(Y^{1:m})\) converges to \(\pi\) in \(W_2\) at rate \(O(\sigma\sqrt{d/m})\) \cite{saremi2023chain}.

To sample from \(\pi\), one first simulates the entire \(M\)-tuple \(Y^{1:M}\) and then returns \(u_M(Y^{1:M})\).  They employ a `Once-At-A-Time' strategy: draw \(Y^1\)  via ULA, then sequentially sample each \(Y^m\mid Y^{1:m-1}\) using MCMC targeting the conditional density of \(Y^m\).  Under mild assumptions this sequence of targets becomes increasingly log-concave in \(m\), but the initial draw of \(Y^1\) can be as challenging as sampling from \(\pi\) itself when \(\sigma\) is large.  Although one can estimate scores via IS or inner-loop posterior MCMC, numerical results show a steep degradation in performance unless \(\sigma\) is carefully tuned—a manifestation of the same “duality of log-concavity” that forces a trade-off between ease of initialization and overall convergence as reported by \citet{grenioux2023onsampling}. 

The time complexity of SMS is $O(NTMn)$ for $N$ final samples, $T$ iterative sampling steps, $n$ MCMC transition steps per sampling step, and $M$ importance samples used for score estimation. 

\subsubsection{Implementation details and hyperparameter settings for the main experiments}\label{app:subsubsec:hyperparameters}

\setlength{\tabcolsep}{2.5pt}
\input{results/tab_our_hyperparam}
\setlength{\tabcolsep}{6pt}

We summarize the hyperparameter values used to tune \ourmethod{} and its variants for the main experiments in \Cref{tab:hyperparam-grid-our}, where these hyperparameters are explained in \Cref{app:subsubsec:method:hyperparam}. 
The choices of score estimator and MCMC kernel types are determined by a coarse grid search using validation metrics; We primarily tune $\nis, \nsteps, \stepsize$ and $\tstart$, as described in \Cref{tab:hyperparam-grid-our}.

\setlength{\tabcolsep}{2.5pt}
\input{results/tab_baseline_hyperparams}
\setlength{\tabcolsep}{6pt}

The hyperparameter search grid for baseline methods is provided in \Cref{tab:baseline-hyperparam}. We largely follow the implementation of the official codebase of \citet{grenioux2024stochastic} (available at \url{https://github.com/h2o64/slips}). For the bi-modal GMM experiments, we tune key hyperparameters within a similar range to those used for \ourmethod{}. For the Rings, Funnel, and Bayesian logistic regression tasks, we adopt the settings reported by \citet{grenioux2024stochastic}.

We use the following target variance information  to initialize AIS, SMC, and SLIPS. Crucially, RDMC and our method do not make use of this  information. 

We restate Assumption A0 from \citet{grenioux2024stochastic}. 
 \begin{assumption}\label{ass:log-concavity}
   (Log-concavity outside a compact). There exist \( R > 0 \) and \( \tau > 0 \) such that \( \pi \) is the convolution of \( \mu \) and \( \mathcal{N}(0, \tau^2 \, \mathbf{I}_d) \), where \( \mu \) is a distribution compactly supported on \( \mathbb{B} = \mathbb{B}(\mathbf{m}_\pi, R ) \), i.e., \( \mu(\mathbb{R}^d \setminus \mathbb{B}) = 0 \).
 \end{assumption}
The values of $R$ and $\tau$ that approximately or exactly satisfy this assumption for different targets are provided in the corresponding experiment subsections.

We next discuss the hyperparameter settings for each of the baseline methods. 

\paragraph{AIS.} AIS uses $T=1,000$ annealing steps. Each transition step uses an MALA kernel with an adaptive step size, initialized at  $\stepsizeinit$, to maintain the acceptance ratio at 75\%. Additionally, MALA runs 4 parallel chains, each with 32 MCMC steps. 
The initial proposal $\rho_0(x)$ is set to $\mathcal{N}(x \mid 0, R^2+\tau^2)$.  

The \textit{only} tunable parameter we consider is $\stepsizeinit$. 

\paragraph{SMC.} SMC follows the same configuration as AIS, except it incorporates ESS-based resampling, introducing the normalized ESS threshold 
$\essthreshold$  as an additional hyperparameter (alongside 
$\stepsizeinit$).

\paragraph{RDMC.} 
RDMC uses $T=1,000$ discretization steps. 
To initialize $\pi_{\tinit}$ in  RDMC, the Langevin-within-Langevin algorithm is simulated using 16 MCMC steps  and 4 MCMC chains. The chains are initialized with an IS approximation of the posterior powered by 128 particles. The initial sample is drawn from $\mathcal{N}(0, (1-\exp\{-2\tinit\}) \mathbb{I}_d\})$ and the initial step size is set to $(1-\exp\{-2\tinit\})/2$

Subsequent steps use a MALA kernel for posterior sampling with adaptive step size to maintain the acceptance ratio at 75\%, 4 parallel chains and 32 MCMC steps. We drop the first 50 MCMC samples to ignore the warm-up period in the estimation.  

The only tunable parameter we consider is $\tinit$. 

\paragraph{SLIPS.} 
SLIPS follows a similar initialization and posterior sampling scheme as RDMC, also with $T=1,000$ discretization steps. However, instead of tuning the value of the initial sampling time $\tinit$, SLIPS makes use of the target variance information and sets $\pi_{\tinit}(x) := \mathcal{N}(x \mid 0, R^2/d + \tau^2)$. 

SLIPS also features a few algorithmic subtleties.
\begin{itemize}
    \item SLIPS uses a stochastic exponential integrator scheme  to simulate from their observational SDE, where ours and RDMC consider the Euler Maruyama scheme. 
    \item  SLIPS discretizes the SDE using evenly spaced points in the log-SNR space, while our method and RDMC perform discretization in the time domain.
    \item  SLIPS reuses the final MCMC samples from the previous step as the initialization for MCMC at the next step, promoting continuity across iterations.
    \item At the final step, SLIPS outputs the estimated denoiser as an approximate sample from the target $\pi$ to align with standard stochastic localization literature. In contrast, our method and RDMC return a random sample generated from the approximated SDE dynamics. 
\end{itemize}

 We consider the following hyperparameters: (i) $\tfinal$ the final integration time of \Cref{eq:slips-reverse-sde}, (ii) $\epsilon$ the initial time to determine the log-SNR-based discretization schedule, and (iii) $\stepsizeinit$, the initial step size for the MALA kernel.

\paragraph{SMS.} We set the number of noisy observations $M$ in SMS to $\lfloor T/2 \rfloor$  where $T=1,000$, and use as many MCMC steps per noise level as SLIPS or
RDMC. This choice of $K$ ensures that the computational complexity of SMS is on par with other baseline methods. 
The Langevin steps are done using ULA as suggested by the authors.

We tune the step size $\stepsize$ and the friction coefficient $\fric$ using recommended values by the authors.

\subsection{Bi-modal gaussian mixtures}\label{app:subsec:gmm}

\paragraph{Target.}
We consider a Gaussian mixture model (GMM) with two components in a $d$-dimensional space. The component means $m_1, m_2$ are randomly initialized from a uniform distribution over a box of width $80$, centered at the origin. Each component has diagonal covariance with all diagonal values equal  to $\sigma^2=2\log 2$. The weights are fixed to $[w_1, w_2]=[0.1,0.9]$ regardless of dimension $d$. Formally, the target density is 
\begin{align}
    \pi(x) &= w_1 \mathcal{N}(x \mid m_1, \sigma^2 \mathbb{I}_d)  +  w_2 \mathcal{N}(x \mid m_2, \sigma^2 \mathbb{I}_d).  
\end{align}

\paragraph{Metrics.}

We compute the estimated weights by assigning samples to different components. 

For a sample $x$, we assign it to the component $k\in \{1,2\}$ with highest posterior probability value, which is computed as 

\[
p(x \textrm{ in component } k  ) = 
\frac{w_k \, \mathcal{N}(x \mid \mu_k, \sigma^2 \mathbb{I}_d)}{\sum_{j=1}^2 w_j \, \mathcal{N}(x \mid \mu_j, \sigma^2 \mathbb{I}_d)}. 
\]

The weight estimates are then formulated as $\hat w_1 = \frac{1}{N} \sum_{i=1}^N\mathbf{1}\{x_i \textrm{ in component } 1\}$ and $\hat w_2 = 1-\hat w_1$. 

The weight estimation ratio bias is computed as 
$|\hat w_1 / (\hat w_1 + \hat w_2) - w_1 / (w_1+w_2)| = |\hat w_1 -w_1|$. 

The log normalization constant  estimation bias is simply $|\hat Z - Z|$ where $\hat Z$ is the estimated value, for method that computes it.

\paragraph{Additional implementation details.}

\Cref{ass:log-concavity} are satisfied for $R=\max (w_1, w_2) \| m_1 - m_2\| = 0.9 \|m_1 - m_2\|$, and $\tau = \sigma$. These values are used to initialize AIS, SMC, SMS, and SLIPS.  

For \ourmethod{} and its variants, we use DSI to estimate the score. The initial proposal distribution in \Cref{alg:annealed-importance-sampler} is set to $\mathcal{N}(\xzero \mid \xt /\alpha_t, \sigma_t^2 / \alpha_t^2)$.

\begin{figure}[!t]
    \centering
\includegraphics[width=0.8\linewidth]{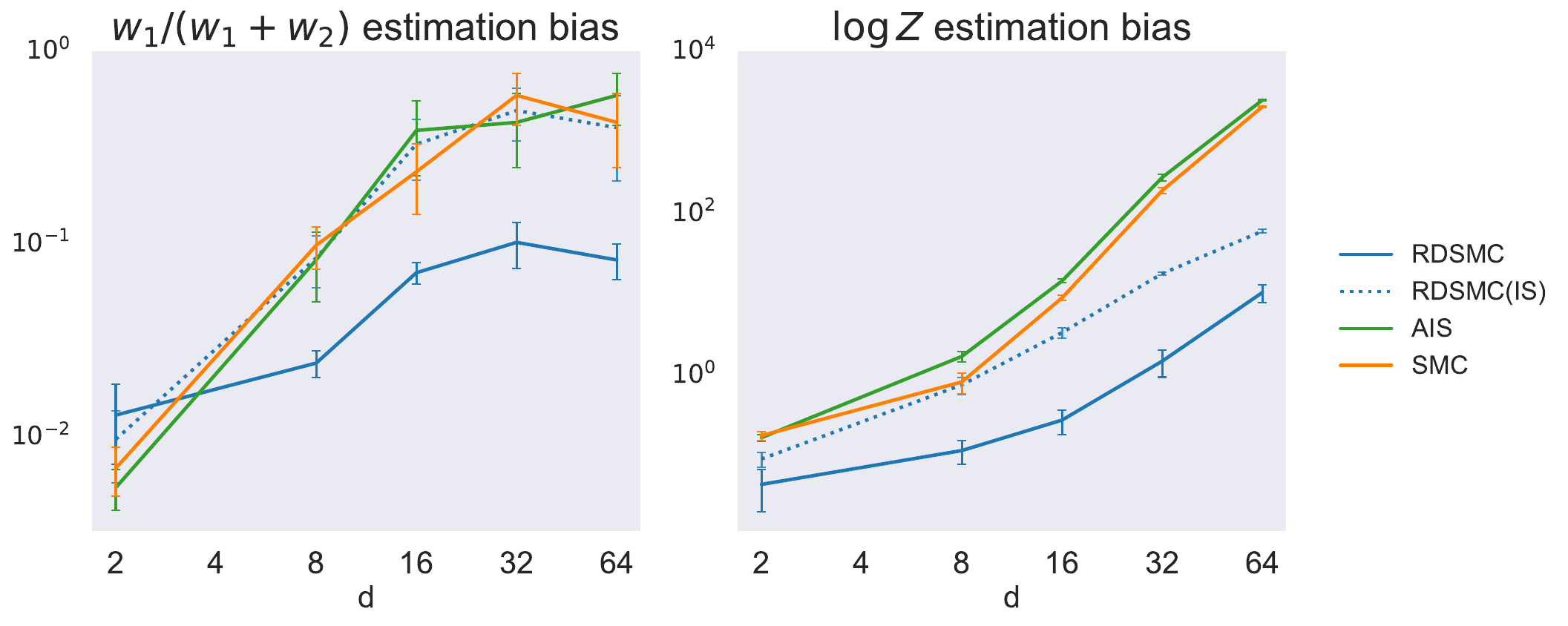}
    \caption{Estimation bias of weight ratio $w_1/(w_1+w_2)$ and log normalization constant $\log Z$, with hyperparameters  selected to minimize the  $\log Z$ estimation bias. We observe that all methods exhibit monotonic trends in both bias metrics.}
    \label{fig:gmm-order-by-logZ}
\end{figure}

\paragraph{Additional results.} 

The hyperparameters used for results in \Cref{subfig:gmm-bias-vs-d} are selected based on minimizing the estimation bias of weight ratio $w_1/(w_1+w_2)$. Here, we also report results using hyperparameters tuned by minimizing the $\log Z$ bias in \Cref{fig:gmm-order-by-logZ}. Results for \ourmethod{} (Proposal), SMS, RDMC, and SLIPS are not included as they do not provide normalization constant estimate. 
The trends are largely similar, except that the $\log Z$ estimation bias curve appears monotonic.

\subsection{Rings and Funnel distributions}
\label{app:subsec:rings-and-funnel}
\paragraph{Targets.}

The Rings distribution \citep{grenioux2024stochastic} is defined via an inverse polar reparameterization of a base distribution $p_z$, which is factorized into two independent univariate marginals $p_r$ and $p_{\theta}$:
\begin{itemize}
    \item $p_r$ is a mixture of four Gaussian distributions, each with mean located at integer radial positions $i+1$ for $i \in \{0, 1, 2, 3\}$ and variance $0.15^2$.
    \item The angular component $p_{\theta}$ is uniformly distributed over $[0, 2\pi]$. 
\end{itemize}

The Funnel distribution \citep{neal2003slice} is characterized by a hierarchical density structure in a 10-dimensional space, where the first dimension $x_1$ is drawn from a Gaussian distribution with zero mean and variance $\sigma^2 = 9$. The subsequent dimensions $x_{2:10}$ are conditionally Gaussian, with variances exponentially dependent on the first dimension. We follow \citet{grenioux2024stochastic} to use the following Funnel density 
\begin{align}
\pi(x_{1:10}) = \mathcal{N}(x_1; 0, \sigma^2) \mathcal{N}(x_{2:10}; 0, \exp(x_1) \mathbb{I}_9).
\end{align}

\paragraph{Metrics.}
The entropic regularized 2-Wasserstein distance between two distributions $\mu$ and $\nu$ is defined by 
\begin{equation}
    \textstyle W_{2,\varepsilon}(\mu, \nu) = \inf\{\int_{\mathbb{R}^d \times \mathbb{R}^d}\|x_1-x_0\|^2\mathrm{d} \pi(x_0,x_1) - H(\pi): \pi_0=\mu, \pi_1=\nu\}^{1/2} \; ,
\end{equation}
where $\varepsilon > 0$ is a regularization hyper-parameter and $H(\pi) = -\int_{\mathbb{R}^d \times \mathbb{R}^d} \log \pi(x_0,x_1) \mathrm{d} \pi( x_0, x_1)$ refers to as the entropy of $\pi$. 
We evaluate the quality of sampling with regularization $\varepsilon=0.05$ via POT library \cite{flamary2021pot}. 

Total Variation Distance (TVD) measures the discrepancy between the true and model-generated distributions by comparing their marginal histograms within a bounded region. 

Let \( \hat{\mu}_i \) and \( \hat{\nu}_i \) represent the probability mass in the \(i\)-th bin for the true and generated distributions, respectively. We  compute the TVD as follows 
\[
\text{TVD} = \frac{1}{2} \sum_{i} |\hat{\mu}_i - \hat{\nu}_i|,
\]

For the radius TVD, we generate the histogram of the radius samples using 256 uniform bins over the interval $[0,8]$. For the angle TVD, we generate the histogram of angle samples  using 256 uniform bins over the interval $[-\pi, \pi]$.

The Kolmogorov-Smirnov distance (KSD) between two distributions $\mu$ and $\nu$ is defined by 
\begin{equation}
    \operatorname{KSD}(\mu,\nu) = \sup_{x \in \mathbb{R}^d} \| F_{\mu}(x) - F_{\nu}(x)\|\;,
\end{equation}
where $F_{\mu}$ and $F_{\nu}$ denotes the cumulative distribution function of $\mu$ and $\nu$ respectively. 
We evaluate the 10-dimensional Funnel samples using the sliced version from \citep[Appendix D.1]{grenioux2023onsampling}.

\paragraph{Additional implementation details.}

For both targets, we use DSI to estimate the score. On Rings, we choose $ R = 4, \tau = 0.15$ that satisfy  \Cref{ass:log-concavity} to initialize AIS, SMC, SMS, and SLIPS; and on Funnel, we use $R = 2.12, \tau = 0.0$,  
following the implementation of \citet{grenioux2024stochastic}.

On Rings, we set the initial proposal distribution 
in \Cref{alg:annealed-importance-sampler} for \ourmethod{} and its variants to $\mathcal{N}(\xzero \mid \xt /\alpha_t, \sigma_t^2 / \alpha_t^2)$. 
On Funnel, the above choice would render numerical instability when $t$ is large. Hence we use $\mathcal{N}(\xzero \mid \xt, \sigma_t^2 / \alpha_t^2)$ as the proposal for the AIS procedure.

\subsection{Bayesian logistic regression}\label{app:subsec:logistic-regression}

\paragraph{Targets.}
We consider four Bayesian logistic regression datasets: Credit and Cancer \citep{nishihara2014parallel}, Ionosphere \citep{sigillito1989classification} and Sonar \citep{gorman1988analysis}.
\begin{itemize}
\item Credit \citep{nishihara2014parallel}: this dataset addresses binary classification of individuals as good or bad credit risks. It contains 1000 data points with 25 standardized features.
\item Cancer \citep{nishihara2014parallel}: this dataset involves classifying recurrence events in breast cancer. It consists of 569 data points in a 31-dimensional standardized feature space.
\item Ionosphere \citep{sigillito1989classification}: this dataset classifies radar signals passing through the ionosphere into good or bad categories. It features 351 data points with dimensionality $d=35$.
\item Sonar \citep{gorman1988analysis}: this dataset differentiates sonar signals reflected from metal cylinders versus cylindrical rocks. It includes 208 data points with $d=61$ standardized features.

\end{itemize}

We consider a training dataset $\mathcal{D} = \{(x_j, y_j)\}_{j=1}^M$ where $x_j \in \mathbb{R}^d$ and $y_j \in \{0, 1\}$ for all $j \in \{1, \ldots, M\}$. 
A Bayesian logistic regression model consists of the following
\begin{align}
p(w, b) &= \mathcal{N}(w; 0, \mathbf{I}_d)\mathcal{N}(b; 0, (2.5)^2) \label{eq:logistic-prior} \\
    p(y \mid x; w, b) &= \text{Bernoulli}(y; \sigma(x^\top w + b))\label{eq:logistic-likelihood}
\end{align}
 where $w \in \mathbb{R}^d$ is a weight vector, $b \in \mathbb{R}$ is an intercept, and $\sigma$ is the sigmoid function.

 The target distribution we are interested in is  the posterior distribution 
\begin{align}
p(w, b \mid \mathcal{D}) \propto p(\mathcal{D} \mid w, b) p(w, b) = \prod_{j=1}^M p(y_j \mid x_j; w, b) p(w, b). \label{eq:logistic-posterior}
\end{align}

\paragraph{Metric.}
Following \citet{grenioux2024stochastic}, we use the following definition of the predictive log-likelihood given a test dataset $\mathcal{D}_{\text{test}}$:
\begin{align}
   \frac{1}{N} \sum_{i=1}^N \sum_{(x,y) \in \mathcal{D}_{\text{test}}} \log p(w_i, b_i) + \log p(y \mid x; w_i, b_i)
\end{align}
where $\{w_i, b_i \}_{i=1}^N$ are posterior samples.

\paragraph{Additional implementation details.}
On all tasks, we use 
$R = 2.5, \tau = 0.0$ in \Cref{ass:log-concavity} to initialize AIS, SMC, SMS, and SLIPS following the implementation of \citet{grenioux2024stochastic}. 

For \ourmethod{} and its variants, we use TSI to estimate the score and clip the norm of the estimated score within 20 to improve numerical stability. We use $\mathcal{N}(\xzero \mid \xt / \alpha_t, \sigma_t^2/ \alpha_t^2)$ as the initial proposal distribution in the AIS-based score estimation procedure.

\subsection{Ablation experiments}\label{app:subsec:ablation}

In our main experiments, baseline methods follow the implementation of \citet{grenioux2024stochastic} which use more discretization steps $T$ compared to our method, and result in different runtimes. Additionally, AIS, SMC and SLIPS use target variance information while \ourmethod{} does not. 

In this section, we conduct ablation studies to control for these factors and evaluate performance under comparable computational budgets. We discuss the hyperparameter settings in \Cref{app:subsubsec:ablation:hyperparam}. We present the results for the targets in the main experiments with the new hyperparameter settings in \Cref{app:subsubsec:ablation:results}. 

In addition, we include an empirical comparison to the Particle Denoising Diffusion Sampler \citep{phillips2024particle} in \Cref{app:subsubsec:ablation:pdds}.

\subsubsection{Hyperparameter settings}\label{app:subsubsec:ablation:hyperparam}
To enable fair comparisons under similar computational budgets, we fix the compute budget for \ourmethod{} and align the hyperparameters of SLIPS, RDMC, and SMS accordingly, since these methods have comparable computational complexity; however, algorithmic differences lead to  variations in runtime. For AIS and SMC, we adjust their hyperparameters to achieve total runtimes comparable to that of \ourmethod{}. 

In addition, only SLIPS and SMS use target scalar variance to initialize their samplers, while other methods including \ourmethod{} do not  access this information. 

We summarize hyperparameter settings for each method below. 

\paragraph{\ourmethod.}
\begin{itemize}
    \item Use $T=100$ uniform discretization timesteps
    \item Use AIS score estimator based on the DSI, and MALA as the MCMC kernel in AIS
    \item Use an adaptive MCMC step size, similar to that used in the baseline SLIPS, AIS, and SMC methods (see description in \Cref{app:subsubsec:method:hyperparam}). This adaptive strategy differs from the fixed step size employed in the main experiments which requires tuning. The initial MCMC step size is 1 for GMM targets and 0.05 for others. 
    \item Use  $M=100$ AIS importance samples 
    \item Use $n=50$ AIS annealing steps 
    \item Use  $m=1$ MCMC transition per AIS step  
    \item Use $\essthreshold=0.3$ for the ESS-based resampling threshold 
    \item The \textit{only} tunable hyperparameter is the starting step for resampling, $\tstart$, which we vary in $\{0,0.1,0.2,\cdots,1.0\}$ unless otherwise specified. Recall that  we  apply resampling at step $t$ when $t \leq \tstart$ and the normalized ESS $< \essthreshold$. 
\end{itemize}

We treat \ourmethod{} (IS) as a special case of RDSMC with $\tstart = 0$, which is included within the generic \ourmethod{} framework as part of hyperparameter tuning. The final hyperparameter values tuned by validation metrics are reported in \Cref{tab:ablation-rdsmc-hyperparams}.

\begin{table}[!t]
    \centering
\input{new_ablation_results/ablation_rdsmc_hyperparams}
    \caption{Ablation experiments: final hyperparameter settings for \ourmethod{}, selected by target-dependent validation metrics. }
    \label{tab:ablation-rdsmc-hyperparams}
\end{table}

\ourmethod{} (Proposal) uses the same configuration as \ourmethod{} except that no resampling or weighting is applied throughout the sampling process.

\begin{table}[!t]
    \centering
    \input{new_ablation_results/tab_ablation_ais_hyperparams}
    \caption{Ablation experiments: final hyperparameter settings for AIS and SMC, tuned to achieve comparable runtime to \ourmethod{} (with $T=100$ time steps, $M=100$ AIS importance samples, and $n=50$ AIS annealing steps) on a single NVIDIA A100 GPU (40 GB memory). While the $M$ importance samples in \ourmethod{} can be processed in parallel, computation can still result in slower performance on a single device. We therefore increase $T$ and $n$ for AIS and SMC to match the runtime of \ourmethod{}.}
    \label{tab:ablation-ais-hyperparams}
\end{table}

\paragraph{AIS and SMC.} To generate $N$ final samples, recall that the computational complexity of AIS and SMC is $\bigO(NTn)$ for $T$ annealing steps and $n$ MCMC transitions per step; in contrast,  \ourmethod{}'s complexity is $\bigO(NTMn)$ for $T$ discretization steps, $M$ AIS importance samples, $n$ AIS annealing steps (with a minor abuse of notation).

While the $M$ AIS importance samples in \ourmethod{} can be processed in parallel, computation can still result in in slower runtime on a single device. Therefore, we adjust the number of annealing steps $T$ and the number of MCMC transitions per step  $n$  to ensure that the total runtime on a single NVIDIA A100 GPU (40 GB memory) is comparable to that of \ourmethod{}. The resulting hyperparameter values are provided in \Cref{tab:ablation-ais-hyperparams}.

Additionally, in contrast to the main experiments following \citet{grenioux2024stochastic}, we do \textit{not} use the target variance to initialize the initial proposal in AIS and SMC. Instead, we use the initial proposal $\mathcal{N}(0,1)$ which coincides with the base distribution of \ourmethod{}. We use the same adaptive MCMC step size strategy as in \ourmethod. For SMC, the ESS-based resampling threshold is set to 0.3. 

Other aspects of AIS and SMC are the same as in the main experiments. 

\paragraph{SLIPS and RDMC.}
We match the computational budget of each score estimation step to that of \ourmethod{}. Specifically, both SLIPS and RDMC use $T=100$ discretization steps; in every  step, they use $M=100$ parallel MCMC chain,  each with $n=50$ transitions, for score estimation. Consequently, the  asymptotic complexity of SLIPS and RDMC (after the initialization stage) is the same as that of  \ourmethod{}, which is  $\bigO(NTMn)$ to produce $N$ final samples.   

However, because SLIPS and RDMC require additional computation for initialization and their score estimation relies on MCMC, whereas \ourmethod{} starts from the base distribution and uses AIS for score estimation, the overall runtimes of these methods differ. 

Other aspects of SLIPS and RDMC are the same as in the main experiments

\paragraph{SMS.} We use $T=100$ steps, $n=50$ MCMC transitions per sampling step  and $M=100$ importance samples for score estimation. Other aspects of SMS are the same  as in the main experiments.

\subsubsection{Ablation experiment results}\label{app:subsubsec:ablation:results}

We revisit the targets studied in the main experiment using the new hyperparameter settings and summarize the results. Each experiment is run with 10 random seeds, and hyperparameters are selected using the corresponding validation metrics.

\paragraph{Runtime.} 
\begin{figure}[!t]
    \centering
    \includegraphics[width=1.0\linewidth]{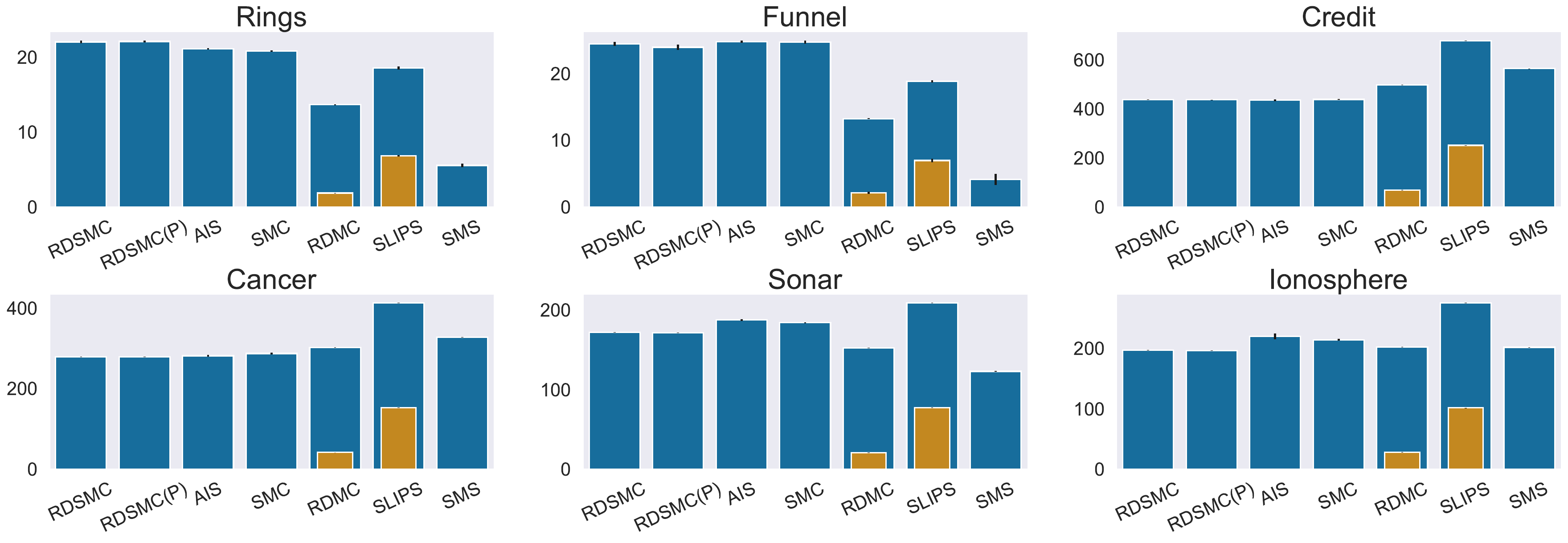}
    \caption{Ablation experiments: runtimes (in seconds) for all methods across different targets. Each blue bar represents the total wall-clock time averaged over 10 random seeds with error bar denoting 2 standard errors, measured on a single NVIDIA A100 GPU (40 GB). For RDMC and SLIPS, we also include the initialization time indicated by the orange bars.  RDSMC(P) denotes \ourmethod{} (Proposal). The GMM results are reported separately in the final panel of \Cref{fig:ablation:gmm}. AIS and SMC are configured to achieve  runtimes comparable to \ourmethod{}. While RDMC, SLIPS and SMS have the same asymptotic complexity of \ourmethod{}, their actual runtimes differ due to algorithmic differences.}
    \label{fig:ablation:runtime}
\end{figure}

\begin{figure}[!t]
    \centering
    \includegraphics[width=1.0\linewidth]{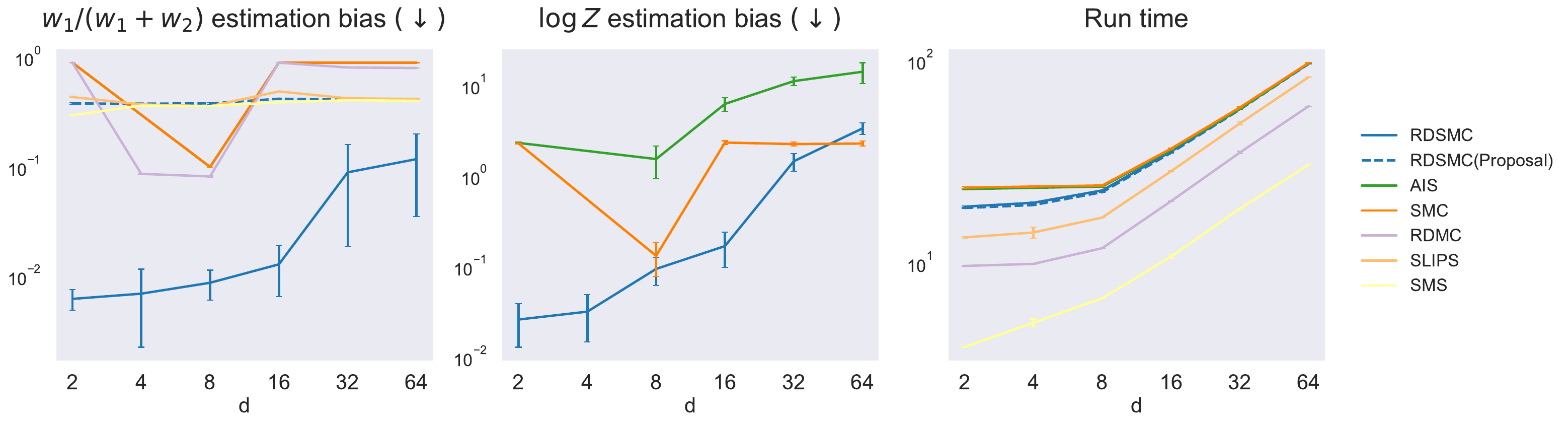}
    \caption{Ablation experiments: estimation bias of weight ratio $w_1/(w_1+w_2)$ (left) and log-normalization constant $\log Z$ (middle), and runtime (in seconds, right) versus dimension $d$. Results are averaged over 10 seeds with error bars showing one standard error. Note that SMS, RDMC, and SLIPS do not provide estimates of $\log Z$. For both metrics, the estimation bias increases with $d$ for all methods, while \ourmethod{} consistently outperforms the baselines. The runtime for \ourmethod{} (including the Proposal variant). SMC and AIS are configured for comparable runtime of \ourmethod{}, whereas SLIPS, RDMC, and SMS exhibit shorter runtimes owing to algorithmic differences, despite having the same asymptotic complexity as \ourmethod{}.}
    \label{fig:ablation:gmm}
\end{figure}

The overall runtimes for all methods across all targets are shown in \Cref{fig:ablation:runtime}, except for the GMM targets, which are presented in the final panel of \Cref{fig:ablation:gmm}. 

We note that AIS and SMC are manually 
configured to achieve runtimes comparable to RDSMC. While RDMC, SLIPS and SMS have the
same asymptotic complexity of RDSMC, their actual runtimes differ due to algorithmic differences. In GMM (with varying $d$), Rings, and Funnel, we observe shorter runtimes for RDMC, SLIPS, and SMS compared to \ourmethod{}. In contrast, for Credit, Cancer, Sonar, and Ionosphere, SLIPS tends to have a longer runtime. Notably, the initialization phase of RDMC and SLIPS accounts for a substantial portion of their total runtime.

\paragraph{Bi-modal GMM.}

We report the results in \Cref{fig:ablation:gmm}. In the first two panels, we observe that for both the weight ratio $w_1 / (w_1 + w_2)$ and the log normalization constant $\log Z$, \ourmethod{} consistently outperforms the other methods. These results are consistent with the findings in the main experiments in \Cref{fig:gmm}, despite a different hyperparameter setting. 
The last panel summarizes the runtime comparison. 

The optimal hyperparameters for each method are selected based on the lowest estimation bias of the weight ratio, consistent with the procedure used in the main experiments.

\paragraph{Rings and Funnel.} 
\begin{table}[!t]
    \centering
\input{new_ablation_results/tab_rings_and_funnel_0916}
    \vspace{0.5em}
    \caption{Ablation experiments: Rings and Funnel targets (mean $\pm$ standard error over 10 seeds). * indicates the best average results and bold indicates 95\% confidence interval overlap with that of the best average result.   AIS and SMC achieve the lowest average Radius TVD and $\log Z$ bias on Rings, and the lowest $\log Z$ bias on Funnel, with \ourmethod{} remaining  comparable when accounting for standard errors. \ourmethod{} achieves the lowest Sliced KSD on Funnel.}
    \label{tab:ablation:rings-and-funnel}
\end{table}

We report  the results in \Cref{tab:ablation:rings-and-funnel}. On synthetic targets, AIS and SMC achieve the lowest average  Radius TVD and $\log Z$ bias on Rings and lowest average $\log Z$ bias on Funnel, while \ourmethod{} achieves comparable results on these metrics and the lowest average Sliced KSD on Funnel.

Note that these observations differ slightly from those in the main experiments (\Cref{tab:rings-and-funnel}) due to changes in the hyperparameter settings. In particular, AIS and SMC no longer use target-variance scaling for initialization, which appears to improve their performance in this case. 

In this new setting, AIS and SMC demonstrate more robust overall performance. While \ourmethod{} remains competitive, it exhibits greater variability in the reported metrics (indicated by the larget standard errors). Nevertheless, \ourmethod{} tends to outperform diffusion-based MC samplers that do not have SMC correction, including its Proposal variant, SLIPS, and RDMC.

The optimal hyperparameters for each method are selected based on the lowest Radius TVD for Rings and lowest Sliced KSD for Funnel, consistent with the procedure used in the main experiments. 

Finally, we note that  for \ourmethod{} on the Rings target,  when the hyperparameter $\tstart > 0.8$, the magnitude of the estimated $\log Z$ becomes significantly larger than when $\tstart \leq 0.8$. For this reason, we tune $\tstart \in \{0, 0.1, \ldots, 0.8\}$, within which the estimated $\log Z$ values remain consistent.

\paragraph{Bayesian logistic regression.}

\begin{table}[!t]
    \centering
     \resizebox{1.0\textwidth}{!}{%
\input{new_ablation_results/tab_ablation_logistic_regression_lppd_0916}
}
    \caption{Ablation experiments: Bayesian logistic regression with test log pointwise predictive density (Test LPPD, with mean $\pm$ standard error) averaged over 10  seeds. * indicates the best average result and bold indicates 95\% confidence interval overlap with that of the best average result.  \ourmethod{}, AIS, SMC, and SLIPS achieve the best overall performance.  \ourmethod{} outperforms RDSMC (Proposal) and RDMC in most cases, highlighting the effectiveness of the SMC correction. 
}
    \label{tab:ablation:logistic:lppd}
\end{table}

Different from the main experiments in \Cref{tab:ablation:logistic:lppd}, we consider a different metric Test LPPD, the log pointwise predictive density on the test set, defined as follows 

\begin{align}
  \text{Test LPPD} 
&= \sum_{(x, y) \in \mathcal{D}_{\textrm{test}}} 
\log \left( 
    \int p(y_i \mid x_i; w,b) \, 
    p(w, b \mid \mathcal{D}) \, \mathrm{d}w,b
\right) 
\label{eq:test-lppd}  
\end{align}
where the likelihood $p(y_i \mid x_i; w,b) $ is defined in \Cref{eq:logistic-likelihood} and the posterior $p(w, b \mid \mathcal{D})$ is defined in \Cref{eq:logistic-posterior}. 

We use final (weighted) samples from each method to approximate the Test LPPD in \Cref{eq:test-lppd}. 

We report the results in \Cref{tab:ablation:logistic:lppd}. We observe that \ourmethod{}, AIS, SMC, and SLIPS achieve the best overall performance.  Compared to SLIPS, \ourmethod{} does not require target-variance-based initialization. Moreover, \ourmethod{} outperforms RDSMC (Proposal) and RDMC in most cases, highlighting the effectiveness of the SMC correction. However, we observe that results for \ourmethod{} exhibit large variance, as the case in the main experiments from \Cref{tab:logistic-regression}. 

The optimal hyperparameters for each method are selected based on the  highest LPPD on a heldout validation set, computed analogously to the Test LPPD. 

\subsubsection{Comparison to Particle Denoising Diffusion Sampler on the Funnel target }\label{app:subsubsec:ablation:pdds}
Finally, we present a comparison to the Particle Denoising Diffusion Sampler (PDDS) \citep{phillips2024particle}, which is also a diffusion-based SMC sampler. However, unlike \ourmethod{}, PDDS requires training.

We evaluate PDDS on the Funnel target using 10 random seeds. Experiments are conducted on an NVIDIA RTX A6000 GPU, whereas our previous experiments use an NVIDIA A100 GPU. In addition, PDDS is implemented in JAX (following the official implementation at \url{https://github.com/angusphillips/particle_denoising_diffusion_sampler#}
), while \ourmethod{} and the other baselines are implemented in PyTorch.

 PDDS requires approximately $82.63 \pm 0.43$ seconds for training and $15.08\pm 0.073$ seconds for final sampling of $4096$ particles, resulting in a total runtime of about $97.70\pm 0.47$ seconds (reported as  mean $\pm$ standard error). The runtimes of our method and other training-free baselines are reported in \Cref{fig:ablation:runtime}, which are all below 25 seconds. However, due to differences in hardware and software frameworks, PDDS's runtime is not directly comparable to those of the previous experiments.

PDDS achieves a  $\log Z$ bias of $0.26 \pm 0.04$ and a mean Sliced KSD of $0.10 \pm 0.00$. Compared with the results in \Cref{tab:ablation:rings-and-funnel}, PDDS exhibits a slightly lower average $\log Z$ bias than \ourmethod{}, but higher than AIS and SMC. Its Sliced KSD is also higher than that of \ourmethod{}, AIS, and SMC.

%% file: results/tab_our_hyperparam.tex
\begin{table}[!ht]
\centering
\begin{tabular}{lcccccc}
\toprule
Target &
$\nis$ & $\nsteps$ & $\stepsize$ & $\tstart$ & score est. & mcmc kernel\\
\midrule
GMM ($d=2,4$) & \{10, 100\} & \{1, 10\} & \{1.0\} & \{0.5, 0.8\} & DSI  & HMC\\
GMM ($d=8\text{ - }64$) & \{10, 100\} & \{ 1, 10, 50\} & \{1.0\}  & \{0.5, 0.8\} & DSI  & HMC\\
Rings & \{100\} & \{1, 10\} & \{0.01, 0.05\} & \{0.5\} & DSI  & MALA \\
Funnel & \{100\} & \{1, 10, 50\} & \{0.01, 0.05\} & \{0.5, 0.8\} & DSI & MALA \\
Logistic Reg. & \{ 10, 100\} & \{10, 50\} & \{0.01, 0.005\} & \{0.5, 0.8\} & TSI & MALA \\ 
\bottomrule
\end{tabular}
\vspace{0.5em}
\caption{Hyperparameter grid used for tuning \ourmethod{} in the main experiments. We fix the number of sampling steps at $T=100$, set $\msteps=1$ MCMC step per AIS transition, and use an ESS threshold of $\essthreshold=0.3$. Score estimator and MCMC kernel types are selected via a coarse grid search using validation metrics. \ourmethod{} (IS) and \ourmethod{} (Proposal) use the same grid with $\essthreshold = 0$, excluding $\tstart$  as it does apply in these settings.}
\label{tab:hyperparam-grid-our}
\end{table}

%% file: results/tab_baseline_hyperparams.tex
\begin{table}[!t]
\centering
\begin{tabular}{l c c c c c}
\toprule
\textbf{Target} 
& \multicolumn{1}{c}{\textbf{AIS}} 
& \multicolumn{2}{c}{\textbf{SMC}} 
& \multicolumn{2}{c}{\textbf{SMS}}  \\
\cmidrule(lr){2-2} 
\cmidrule(lr){3-4} 
\cmidrule(lr){5-6}
& $\stepsizeinit$ 
& $\stepsizeinit$ & $\essthreshold$ 
& $\stepsize$ & $\fric$
 \\
\midrule
GMM   & \{0.1, 0.5, 1.0\} &  \{0.1, 0.5, 1.0\} & \{0.3, 1.0\} & \{0.03, 1.0\} & \{0.0625, 0.05\}  \\
Rings & \{0.01\} & \{0.01\} & \{0.3, 1.0\} & \{0.03, 1.0\} &  \{0.0625, 0.05\}  \\
Funnel & \{0.01\} & \{0.01\} & \{0.3, 1.0\} & \{0.03, 1.0\} & \{0.0625, 0.05\}  \\
Logistic Reg. & \{0.01\} & \{0.01\} & \{0.3, 1.0\} & \{0.03, 1.0\} &  \{0.0625, 0.05\}  \\
\bottomrule
\end{tabular}

\vspace{1em}

\begin{tabular}{l  c c c c}
\toprule
\textbf{Target} 
& \multicolumn{1}{c}{\textbf{RDMC}} 
& \multicolumn{3}{c}{\textbf{SLIPS}} \\
\cmidrule(lr){2-2} 
\cmidrule(lr){3-5} 
& $\exp\{\tinit\}$ 
& $\tfinal$ & $\epsilon$ & $\stepsizeinit$ \\
\midrule
GMM  & \{$ 0.95,  0.9,  0.8, 0.7$ \} & \{150, 300\} & \{0.03, 0.05, 0.1, 0.2, 0.4\} & \{0.1, 0.5, 1.0\} \\
Rings &  \{$ 0.95,  0.9,  0.8, 0.7$ \}  & \{150, 300\} & \{0.1, 0.2, 0.4, 1.0, 1.2\} & \{1e-5\} \\
Funnel &  \{$ 0.95,  0.9,  0.8, 0.7$ \}  & \{150, 300\} & \{0.1, 0.2, 0.4, 1.0, 1.2\} & \{1e-5\} \\
Logistic  Reg. & \{$ 0.95,  0.9,  0.8, 0.7$ \} & \{150\} & \{0.03, 0.05, 0.1, 0.2, 0.4\} & \{1e-5\}  \\
\bottomrule
\end{tabular}
\vspace{0.5em}
\caption{Hyperparameter grid used for tuning baseline methods in the main experiments. AIS, SMC, RDMC, and SLIPS use $T=1,000$ sampling steps, and SMS uses $K=500$ noisy observations.} 
\label{tab:baseline-hyperparam}
\end{table}

%% file: new_ablation_results/ablation_rdsmc_hyperparams.tex
\begin{tabular}{lc}
\toprule
Target & start resampling time ($\tstart$) \\
\midrule
GMM $(d=2)$       & 0.2 \\
GMM $(d=4)$        & 0.6 \\
GMM $(d=8)$         & 0.4 \\
GMM $(d=16)$        & 0.2 \\
GMM $(d=32)$        & 0.1 \\
GMM $(d=64)$        & 0.4 \\
Rings        & 0.1 \\
Funnel       & 1.0 \\
Credit       & 0.9 \\
Cancer       & 0.6 \\
Ionosphere   & 0.7 \\
Sonar        & 1.0 \\
\bottomrule
\end{tabular}

%% file: new_ablation_results/tab_ablation_ais_hyperparams.tex
\begin{tabular}{lcc}
\toprule
Target & \# of annealing steps ($T$) & \# of MCMC transitions per  step ($n$) \\
\midrule
GMM ($d=2,4,8$)        & 100  & 100 \\
GMM ($d=16$)       & 150  & 100 \\
GMM ($d=32$)       & 240  & 100 \\
GMM ($d=64$)       & 400  & 100 \\
Rings        & 100  & 70  \\
Funnel       & 100  & 100 \\
Credit       & 1000 & 140 \\
Cancer       & 1000 & 90  \\
Ionosphere   & 1000 & 70  \\
Sonar        & 1000 & 60  \\
\bottomrule
\end{tabular}

%% file: new_ablation_results/tab_rings_and_funnel_0916.tex
\begin{tabular}{lcccc}
\hline
\multirow{2}{*}{Algorithm} & \multicolumn{2}{c}{Rings} & \multicolumn{2}{c}{Funnel} \\
\cmidrule(r{.5em}){2-3}\cmidrule(l{.5em}){4-5}
& Radius TVD $\downarrow$ & log Z bias $\downarrow$ & Sliced KSD $\downarrow$ & log Z bias $\downarrow$ \\
\hline
RDSMC & 0.118 $\pm$ 0.002 & \textbf{0.013 $\pm$ 0.004} & \textbf{0.075 $\pm$ 0.005}* & \textbf{0.286 $\pm$ 0.068} \\
RDSMC(Proposal) & \textbf{0.091 $\pm$ 0.003} & N/A & 0.327 $\pm$ 0.033 & N/A \\
RDMC & 0.440 $\pm$ 0.003 & N/A & 0.129 $\pm$ 0.000 & N/A \\
SLIPS & 0.279 $\pm$ 0.003 & N/A & \textbf{0.076 $\pm$ 0.001} & N/A \\
AIS & \textbf{0.088 $\pm$ 0.003} & \textbf{0.004 $\pm$ 0.002}* & \textbf{0.087 $\pm$ 0.002} & \textbf{0.173 $\pm$ 0.010} \\
SMC & \textbf{0.088 $\pm$ 0.002}* & \textbf{0.006 $\pm$ 0.002} & \textbf{0.082 $\pm$ 0.004} & \textbf{0.169 $\pm$ 0.012}* \\
SMS & 0.282 $\pm$ 0.010 & N/A & 0.161 $\pm$ 0.000 & N/A \\
\hline
\end{tabular}

%% file: new_ablation_results/tab_ablation_logistic_regression_lppd_0916.tex
\begin{tabular}{lcccc}
\hline
Test LPPD $\uparrow$ & Credit ($d=25$) & Cancer ($d=31$) & Ionosphere ($d=35$) & Sonar ($d=61$) \\
\hline
RDSMC & \textbf{-94.54 $\pm$ 1.34} & \textbf{-10.59 $\pm$ 0.45} & \textbf{-25.97 $\pm$ 0.73} & \textbf{-18.54 $\pm$ 0.28} \\
RDSMC(Proposal) & -94.62 $\pm$ 0.06 & -50.24 $\pm$ 0.16 & \textbf{-24.87 $\pm$ 0.02}* & -18.91 $\pm$ 0.01 \\
RDMC & -138.22 $\pm$ 0.35 & -78.03 $\pm$ 0.12 & -44.16 $\pm$ 0.07 & -28.64 $\pm$ 0.03 \\
SLIPS & \textbf{-92.42 $\pm$ 0.02}* & -10.41 $\pm$ 0.01 & -25.22 $\pm$ 0.01 & \textbf{-18.40 $\pm$ 0.01}* \\
AIS & \textbf{-92.91 $\pm$ 0.44} & \textbf{-10.13 $\pm$ 0.07} & -25.09 $\pm$ 0.02 & \textbf{-18.41 $\pm$ 0.01} \\
SMC & \textbf{-92.55 $\pm$ 0.05} & \textbf{-10.13 $\pm$ 0.01}* & -25.09 $\pm$ 0.02 & \textbf{-18.41 $\pm$ 0.01} \\
SMS & -98.00 $\pm$ 0.27 & -20.47 $\pm$ 0.16 & -26.17 $\pm$ 0.10 & -23.29 $\pm$ 0.08 \\
\hline
\end{tabular}